\documentclass[aps, prd,twocolumn, showpacs, superscriptaddress, groupedaddress]{revtex4} 
\usepackage{graphicx}	
\usepackage{amssymb}
\usepackage{amsmath}
\usepackage{dcolumn}
\usepackage{color}
\usepackage{subfigure, rotating, bm, array}
\usepackage[pagebackref=false, colorlinks=true]{hyperref}
\hypersetup{
linkcolor=blue,     
citecolor=blue,     
urlcolor=blue}      
\begin{document}
\title{Precession of timelike bound orbits in Kerr spacetime}
\author{Parth Bambhaniya}
\email{grcollapse@gmail.com}
\affiliation{International Center for Cosmology, Charusat University, Anand, GUJ 388421, India}
\author{Divyesh N. Solanki}
\email{divyeshsolanki98@gmail.com}
\affiliation{Sardar Vallabhbhai National Institute of Technology, Surat GUJ 395007,  India}
\author{Dipanjan Dey}
\email{dipanjandey.icc@charusat.ac.in}
\affiliation{International Center for Cosmology, Charusat University, Anand, GUJ 388421, India}
\author{Ashok B. Joshi}
\email{gen.rel.joshi@gmail.com}
\affiliation{International Center for Cosmology, Charusat University, Anand, GUJ 388421,  India}
\author{Pankaj S. Joshi}
\email{psjprovost@charusat.ac.in}
\affiliation{International Center for Cosmology, Charusat University, Anand, GUJ 388421, India}
\author{Vishwa Patel}
\email{vishwapatel2550@gmail.com}
\affiliation{PDPIAS, Charusat University, Anand, GUJ 388421, India}

\date{\today}

\begin{abstract}
Astrometric observations of S-stars provide a unique opportunity to probe the nature of Sagittarius-A* (Sgr-A*). 
In view of this, it has become important to understand the nature and behavior of timelike bound trajectories of particles around a massive central object. It is known now that whereas 
the Schwarzschild black hole does not allow the negative precession for the S-stars, the naked singularity spacetimes can admit the positive as well as negative precession for the bound timelike orbits. In this context, we study the perihelion precession of a test particle in the Kerr spacetime geometry. Considering some approximations, we investigate whether the timelike bound orbits of a test particle in Kerr spacetime can have negative precession. In this paper, we only consider low eccentric timelike equatorial orbits. With these considerations, we find that in Kerr spacetimes, negative precession of timelike bound orbits is not allowed.  

\end{abstract}

\maketitle

\section{Introduction}

In recent days, the study of the nature of Sgr-A* has become a subject of great interest. Sgr-A* is a highly dense compact object that exists at the center of our Milky-way galaxy. It is expected that the mass of Sgr-A* is about $4.3\times10^6 M_{\odot}$,  which is located at a distance of $8.2$ kpc ($1~ $kpc $\sim$ $3\times 10^{16}$ km) from the Earth. There are many `S' stars (e.g. S-2, S-38, S-102, etc.) which are orbiting around the compact radio source Sgr-A*. The observational data of S-stars' orbits can reveal the great mystery about the nature of the compact object Sgr-A*.  In order to observe the relativistic corrections in the orbital motion of the S-stars around Sgr-A*,
the GRAVITY and  SINFONI collaborations are monitoring their orbital dynamics continuously.

It is generally believed that super massive black holes (SMBHs) with masses between $10^6-10^9 M_{\odot}$ exist at the centers of most galaxies \cite{Kormendy}. In order to predict the causal nature of the galactic centers, it is very important to investigate different possible observational features, such as the shadows of black holes, the relativistic orbits of the S-stars, accretion disk properties of the compact objects, etc. The Event Horizon Telescope (EHT) collaboration has released the first-ever image of the shadow of a black hole located at the center of the Messier 87 (M87) galaxy \cite{Akiyama:2019fyp}. There are several literature where the detailed analysis of the shadows cast by spherically symmetric and static black holes and other compact objects (e.g. different naked singularities, gravastar, etc.) are studied \cite{Shaikh:2018lcc,Gyulchev:2019tvk, Shaikh2, Nakao1, DipanjanRaji, AJoshi1}. It has been shown that compact objects, like naked singularities can cast similar type of shadow which is expected to be cast by a black hole \cite{Shaikh:2018lcc, DipanjanRaji, AJoshi1}.  Recently, GRAVITY collaboration has shown the possibility of the existence of general relativistic precession of the orbit of S2 star, where they consider Schwarzschild black hole at the center \cite{Gravity}. In our earlier work \cite{Dipanjan}, we have predicted the precession angle of the S2 star considering Schwarzschild black hole and naked singularity at the center. The study of the relativistic orbits of S-stars near the Milky-way galactic center is very important in astrophysics to verify the prediction of the general theory of relativity. There are several works where timelike and lightlike geodesics in different spacetimes are investigated which are very important in the context of recent observations of EHT, GRAVITY and SINFONI collaborations \cite{parsa,Banik:2016qvf,Kovacs:2010xm,Dey:2013yga,Dey+15,Zhou:2014jja,Levin:2009sk,levin,Glampedakis:2002ya,Chu:2017gvt,Dokuchaev:2015zxe,Borka:2012tj,Martinez:2019nor,Fujita:2009bp,Wang:2019rvq,Suzuki:1997by,Zhang:2018eau,Pugliese:2013zma,Farina:1993xw,Dasgupta:2012zf,Shoom,M87,Eisenhauer:2005cv,Eva2,Bhatt1, Pugliese:2011xn, Pugliese:2010he,Hackmann:2015vla,Deng:2020yfm,Gao:2020wjz,Takamori:2020ntj,Collodel:2017end,Ulbricht:2015vwa,Bini:2013tia,Sandoval:2013fua,Thornburg:2016msc}.

The Schwarzschild black hole is a spherically symmetric, static and vacuum solution of the Einstein field equation. It describes spacetime geometry of a non-rotating, uncharged black hole which is characterized by a single parameter, the Schwarzschild mass M of the black hole. In \cite{Parth,Joshi:2019rdo}, we show that the bound orbits of a freely falling particle in Schwarzschild spacetime always precess in the positive direction. In positive precession, the angular distance travelled by a particle between two successive perihelion points is greater than $2\pi$. On the other hand, the precession is called  negative when the angular distance travelled by a particle between two successive perihelion points is less than $2\pi$. In positive precession, it can be shown that  the orbit of a particle always precesses in the direction of the particle motion, whereas in the negative precession, the orbit precesses in the opposite direction of the particle motion. However, the negative precession is forbidden in Schwarzschild spacetime \cite{Parth}. Therefore,  if we consider Sgr-A* to be a Schwarzschild black hole then the observed precession of `S' stars should always be positive. It is generally believed that an extended matter distribution (e.g. clusters of stars, baryonic matter, dark matter, etc. ) can cause or allow for a negative precession of the stars. In \cite{Parth,Joshi:2019rdo}, we study the bound orbits of a test particle in Schwarzschild, Joshi-Malafarina-Narayan (JMN), and Janis-Newman-Winicour (JNW) spacetimes. Both the JMN and JNW spacetimes are naked singularity spacetimes which have spherically symmetric matter distributions \cite{psJoshi1, psJoshi2, Virbhadra:1998dy, JNW}. The JMN and JNW naked singularity spacetimes allow both the negative and positive precession of timelike bound orbits. In \cite{DipanjanRaji}, it is shown that for some parameters' values, when negative precession is allowed in JMN and JNW spacetimes, these spacetimes would not  cast any shadow for the central object. However, they can cast a shadow when the positive precession is allowed in these spacetimes. Therefore, in that earlier work \cite{DipanjanRaji}, we construct a spacetime configuration in which both the negative precession and shadow can exist simultaneously. As we know, the Schwarzschild, Kerr and other black hole solutions can cast a shadow. Therefore, if the negative precession of the orbit of a test particle in other black holes spacetimes is possible then that would be another example along with the spacetime configuration of \cite{DipanjanRaji}, where a negative precession and shadow both can exist simultaneously. In this context, in this paper, we investigate the possibility of a likely negative precession of bound timelike orbits in the Kerr spacetime, which is physically more realistic solution of a black hole as compared to the Schwarzschild black hole solution, which allows for no rotation.

As we know, every celestial body has its own intrinsic spin angular momentum. Therefore, inclusion of a non-zero spin in a non-rotating spacetime makes the modified spacetime  physically more realistic. There are no restrictions on the value of the spin angular momentum of a compact object, as long as it is not a black hole. The spin of a celestial object is typically  represented by a dimensionless spin parameter $\tilde{a}$. This spin parameter can be defined as $\tilde{a}=\frac{c J}{GM^2}$, where $c, J, G, M$ are the velocity of light, intrinsic spin angular momentum of the body, Newton's gravitational constant, and the mass of the celestial body respectively. Earth, with its spin angular momentum $J\sim 7.2\times 10^{33} kg~ m^2 s^{-1}$ and mass $M= 5.972\times 10^{24} kg$ has the spin parameter $\tilde{a}=907$, whereas the sun has spin parameter $\tilde{a}=0.216$ \cite{Nielsen:2016kyw}. On the other hand, a rapidly spinning massive star VFTS102 has $\tilde{a}=75$ \cite{Nielsen:2016kyw}.
Therefore, one can see that the value of the spin parameter of a celestial object can be  much greater than one. However, if we consider Kerr black hole which is a rotating generalization of the Schwarzschild black hole, the spin parameter $\tilde{a}$ cannot be allowed to be greater than unity in order to ensure the existence of an event horizon. The Kerr black hole is a vacuum axi-symmetric solution of Einstein field equations, and it is characterized by two parameters, the total mass of the black hole $M$ and total angular momentum $J$. Kerr spacetime describes a rotating black hole, if $\tilde{a} \leq 1$, whereas it describes a rotating naked singularity spacetime if $\tilde{a} > 1$.  Astrophysically relevant bound trajectories around a Kerr black hole are studied in \cite{rana}. Detailed study of bound nonspacelike geodesics in the Kerr metric is given in \cite{Wilkins, thorne}.

As we mentioned previously, in this paper, our prime focus is to investigate the nature of perihelion precession of timelike bound orbits in Kerr spacetime. In \cite{Dipanjan}, we show that Schwarzschild spacetime  does not admit any negative precession. In this paper, we show that the precession of the timelike bound orbits in Kerr spacetime is also always positive, when we consider only low eccentric, equatorial, bound orbits. With the small eccentricity approximation, one can analyse the nature of the timelike bound orbits in the strong field region of Kerr spacetime where the relativistic effects on the bound orbits are likely to be observed. 

This paper is organized as follows. In section (\ref{two}), we review the basic properties of the Kerr spacetime and derive the  analytic  solution of the orbit equation for a test particle in the equatorial plane. We numerically solve the orbit equation and show the particle trajectories. In section (\ref{three}), we use an approximate solution of  orbit equation and investigate the nature of perihelion precession of the timelike bound orbits in Kerr spacetime. We summarize and conclude our results in section (\ref{four}). Throughout the paper, we use geometrical units with $G = c = 1$.


\begin{figure*}
	\includegraphics[scale=0.42]{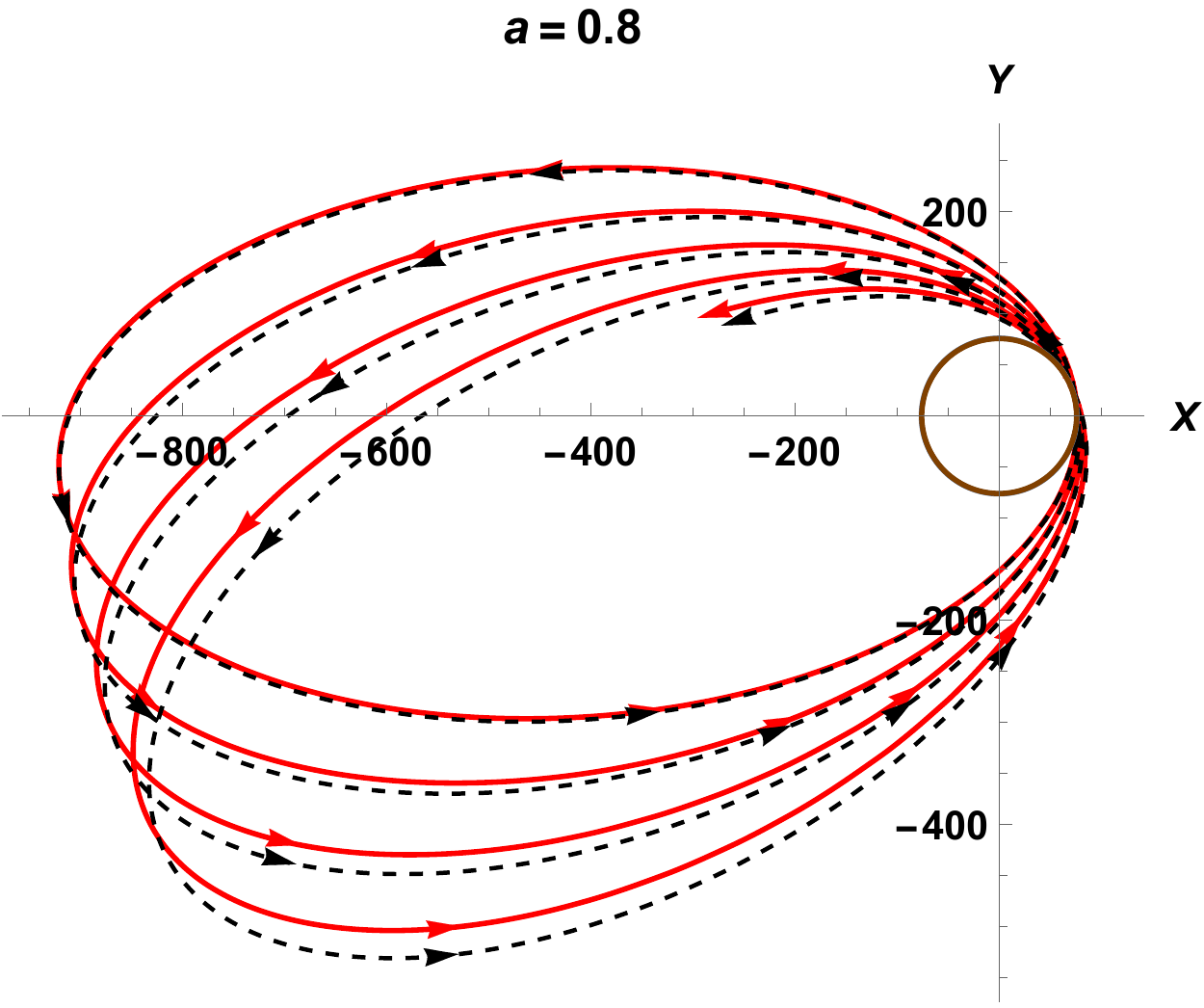}
	\includegraphics[scale=0.42]{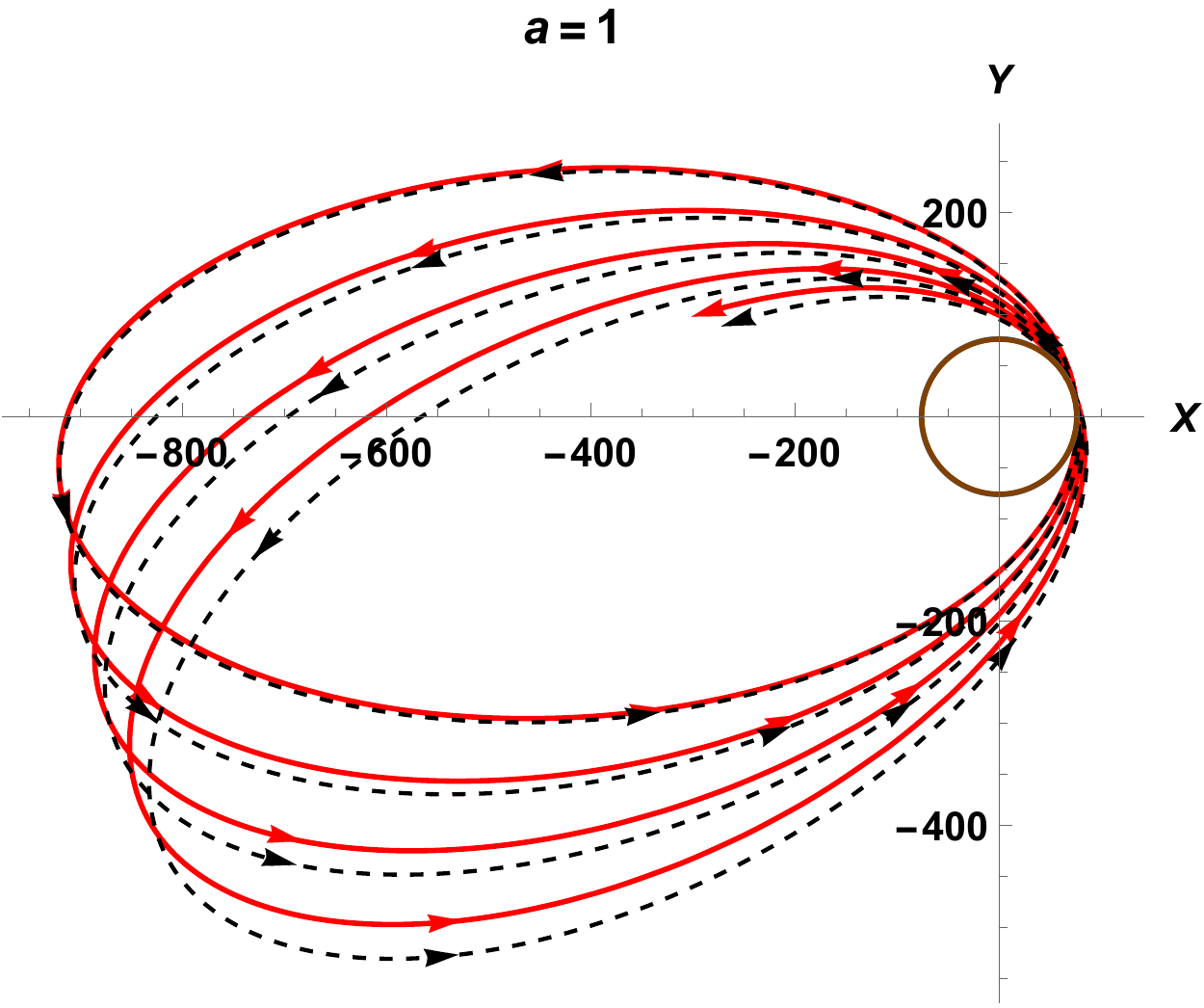}
	\includegraphics[scale=0.42]{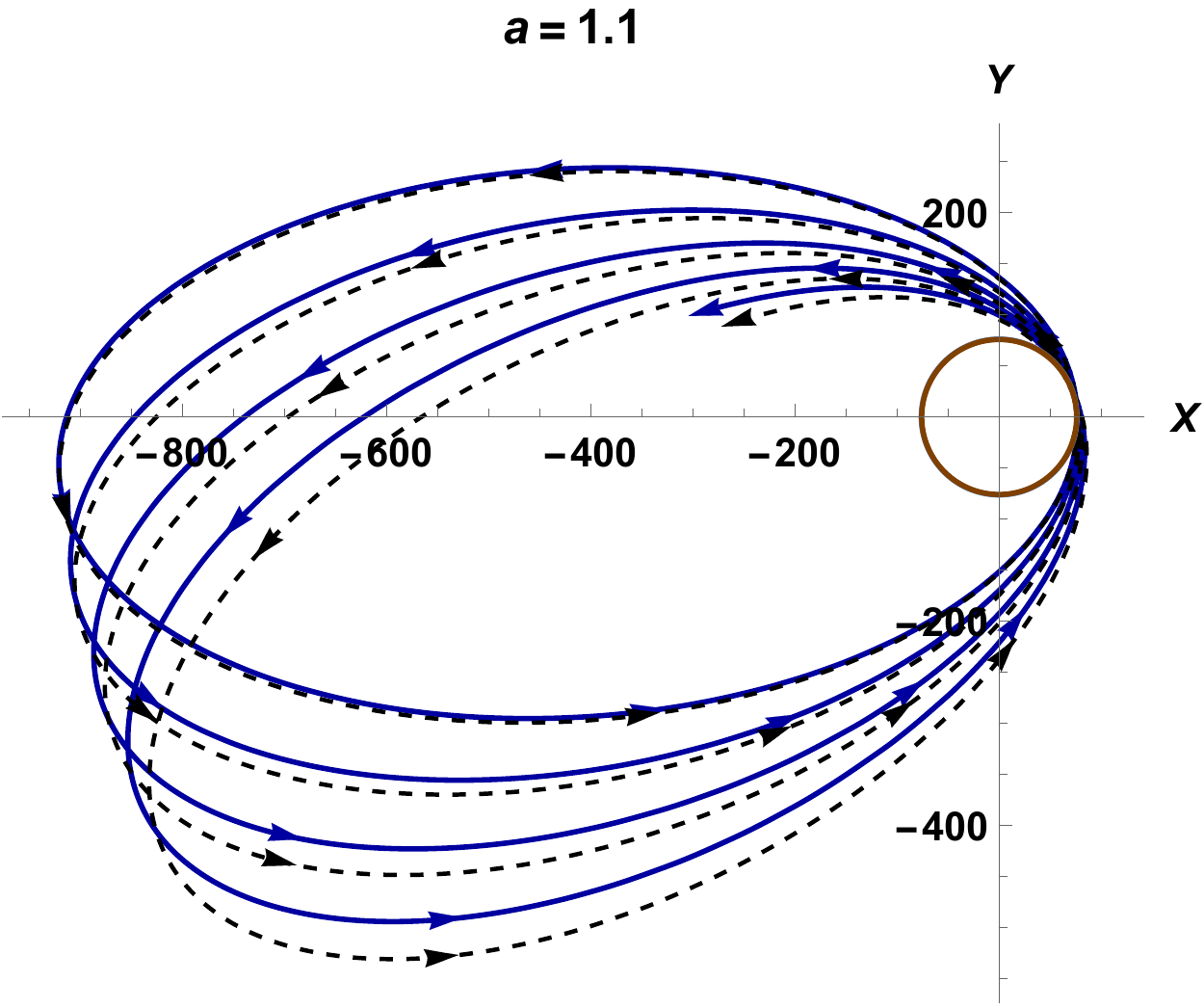}
	\includegraphics[scale=0.42]{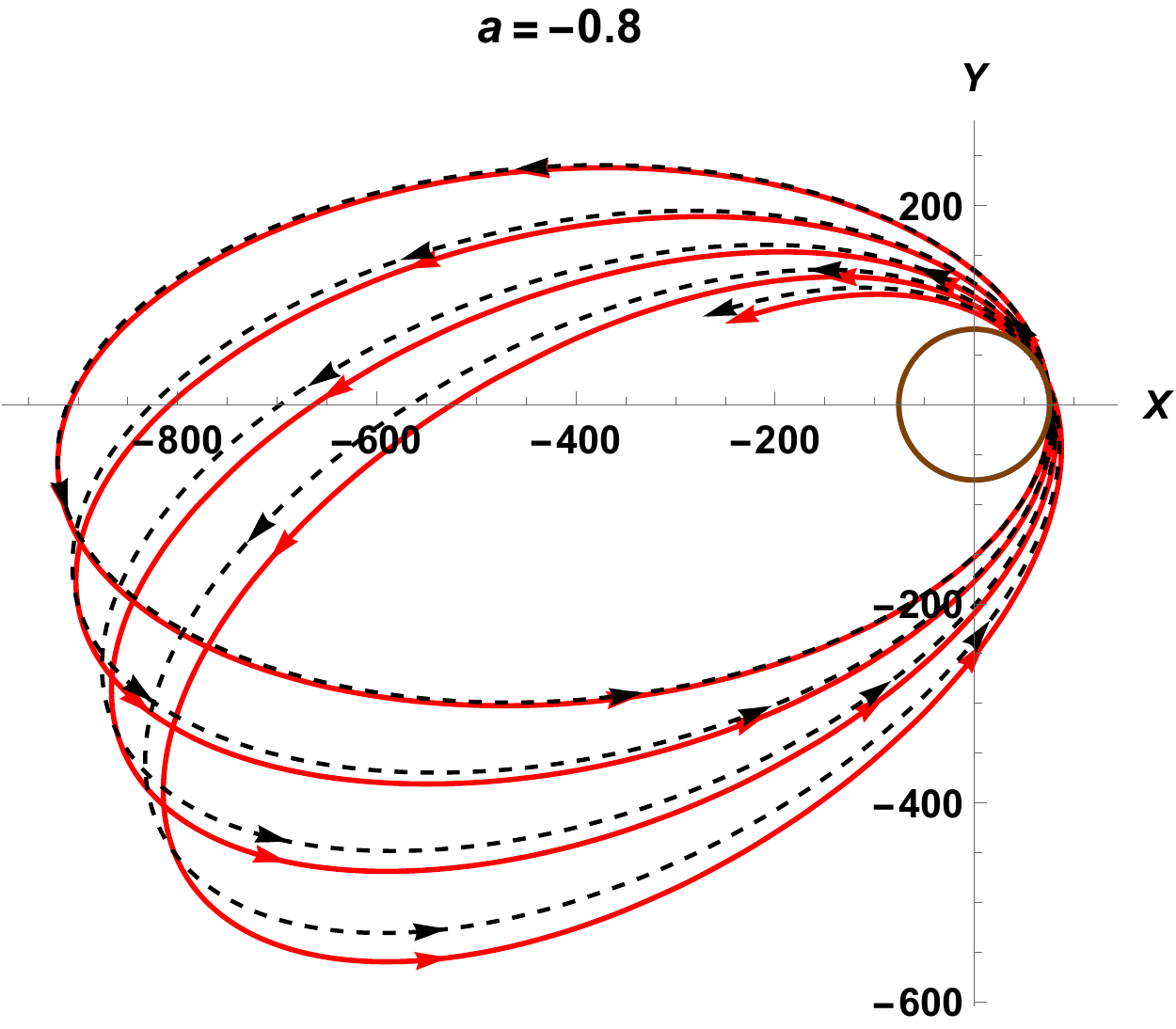}
	\includegraphics[scale=0.42]{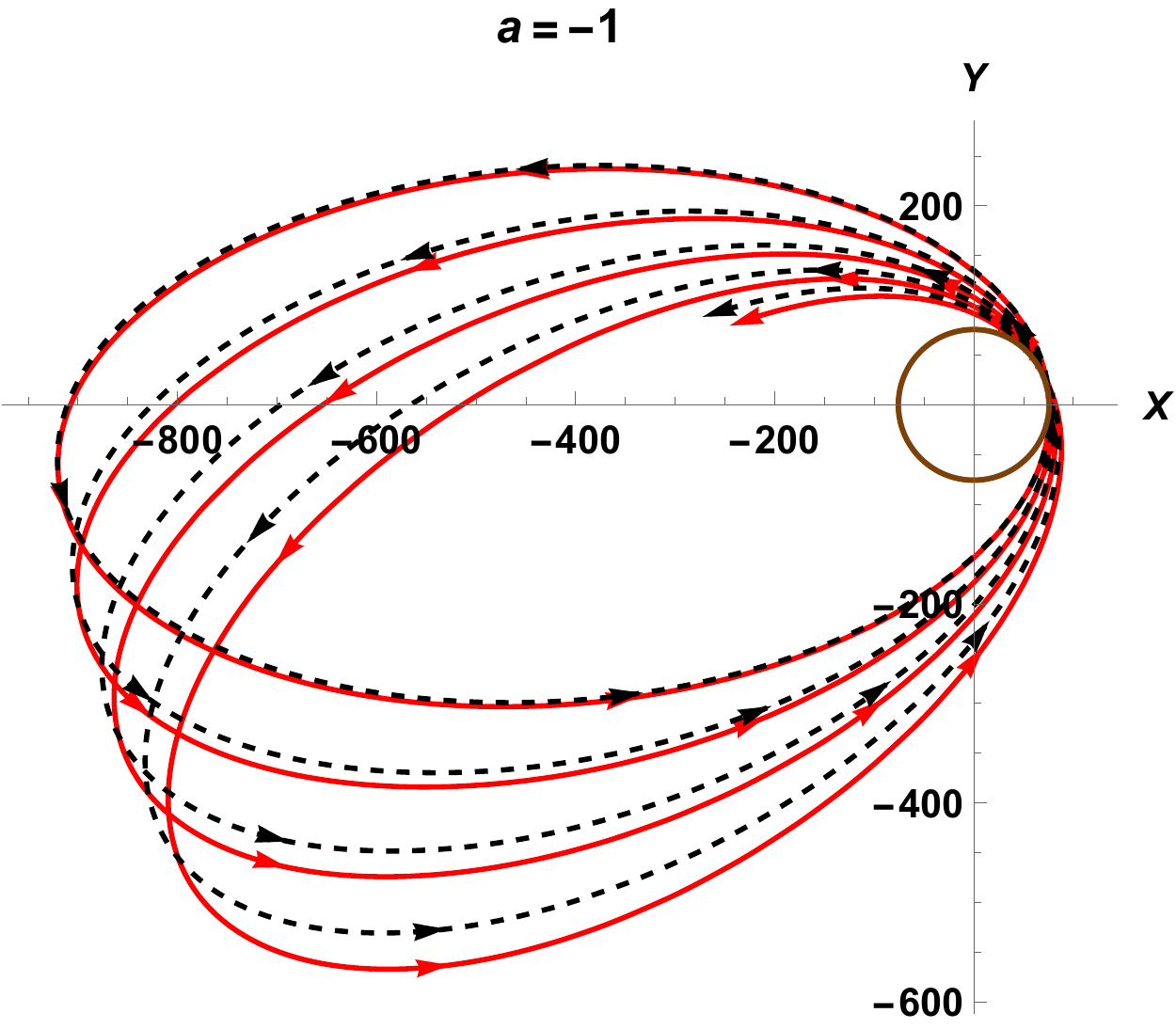}
	\includegraphics[scale=0.42]{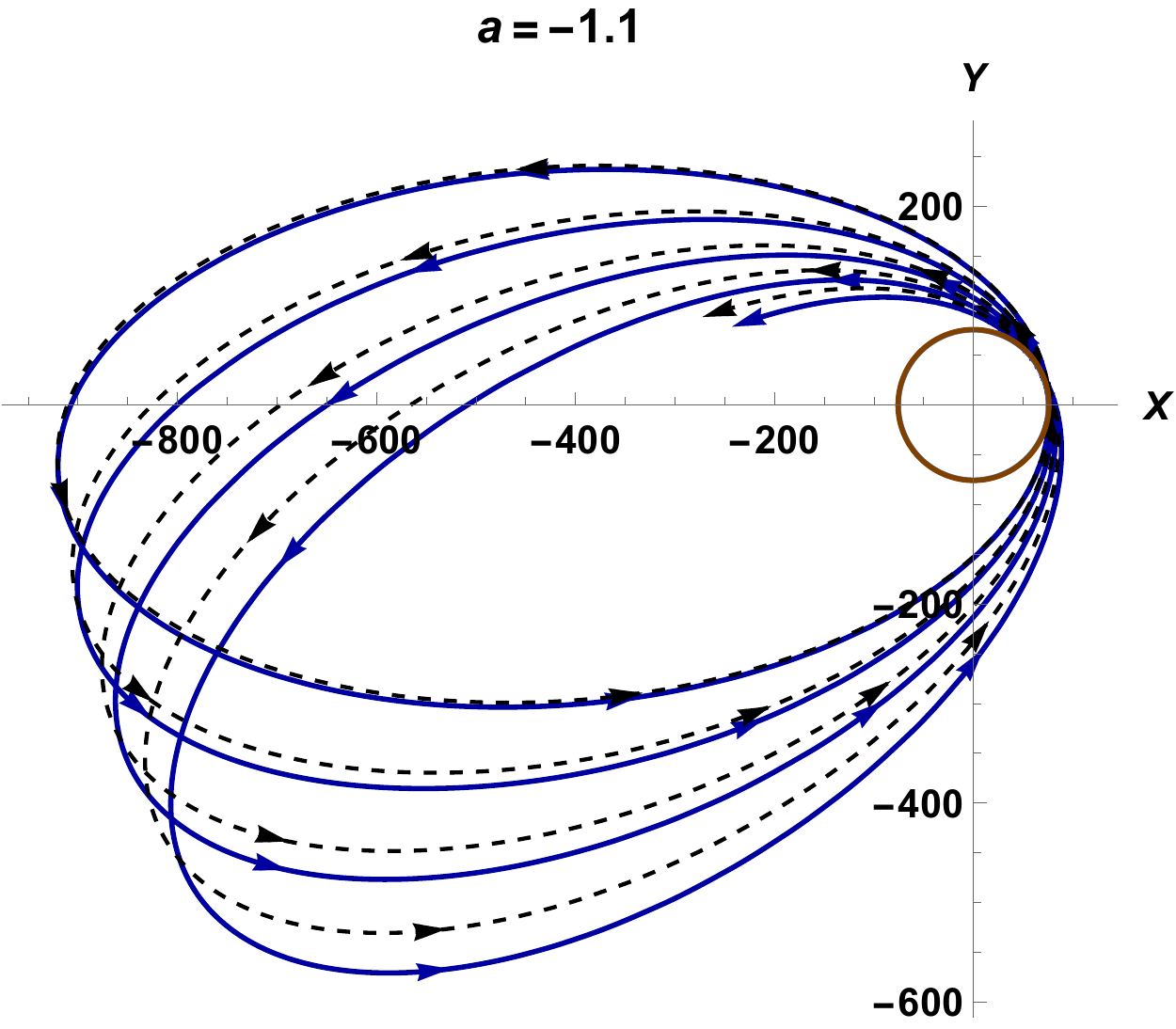}
		\caption{ Above figures show orbits of a test particle in the Kerr spacetime. Red solid lines indicate particle's orbits in the Kerr black hole for $a=0.8, r_{min}=76.15$ and $a=-0.8, r_{min}=75.53$ (first column), in the  extreme Kerr black hole for $a=1, r_{min}=76.22$ and $a=-1, r_{min}=75.45$ (second column) and blue solid lines indicate particle's orbits in the Kerr naked singularity for $a=1.1, r_{min}=76.25$ and $a=-1.1, r_{min}=75.41$ (third column). Black dotted lines indicate particle's orbits in the Schwarzschild spacetime (for a=0, $r_{min}=75.85$). Brown circles represent the minimum approach of the particle towards the center. Here, we considered $M=1$, $L=12$ and total energy $E=-0.001$.}
	\label{orbit1}
    \end{figure*}

\section{Timelike geodesics in Kerr spacetime}
\label{two}
In this section, we study the motion of a test particle in the Kerr spacetime. The stationary, axisymmetric and rotating Kerr spacetime is given in Boyer-Lindquist coordinates as,
\begin{widetext}
\begin{equation}
    ds^2 = -\left(1-\frac{r_s r}{\Sigma}\right)  dt^2 + \frac{\Sigma}{\Delta} dr^2 + \Sigma d\theta^2 + \left(r^2 + a^2 + \frac{r_s r a^2 \sin^2\theta}{\Sigma}\right) \sin^2\theta d\phi^2 - \frac{2 r_s r a \sin^2\theta}{\Sigma}  dt d\phi\,\, , 
    \label{kerr}
\end{equation}
\end{widetext}
where $\Sigma = r^2 + a^2 \cos^2\theta$, 
    $\Delta = r^2 + a^2 - r_s r$ and $r_s= 2 M$. The spin parameter $a$ with a length dimension is related with the total angular momentum $J$ as $a = J/M $. Therefore, in terms of dimensionless spin parameter $\tilde{a}$, we can write $a$ as, $a=M\tilde{a}$. The horizons are defined by the relation $g_{rr}\rightarrow\infty$ which implies that the solution of $\Delta = 0$ can give us the position of the event horizon. Therefore,
    \begin{equation}
        r_\pm = M \pm \sqrt{M^2 - a^2}\,\,.
        \label{horizon}
    \end{equation}
There are two horizons, the event horizon at $r=r_+$ and Cauchy horizon at $r=r_-$. For $a = M$, both these horizons coincide at $r=M$, and these are known as extreme Kerr black holes. When $a > M$, there is no horizon and the Kerr black hole becomes a timelike naked singularity. The study of test particles motion in Kerr spacetime is important to understand the physical processes occuring in these spacetimes and their observational consequences. The Kerr spacetime is independent of $t$ and $\phi$, therefore, the conserved energy ($e$) and the angular momentum ($L$) per unit rest mass are given by,
\begin{equation}
    e = g_{tt} U^t + g_{t\phi} U^{\phi}\,\, ,
    \label{conserved1}
\end{equation}
\begin{equation}
     L = -g_{t\phi} U^t + g_{\phi\phi} U^{\phi}\,\, ,
     \label{conserved2}
\end{equation}
where $U^\mu$ are the components of the four velocity of a test particle and $g_{tt} = \left(1-\frac{r_s r}{\Sigma}\right) $, $g_{rr} = \frac{\Sigma}{\Delta}$, $g_{\theta\theta} = \Sigma$, $g_{\phi\phi} = \left(r^2 + a^2 + \frac{r_s r a^2 \sin^2\theta}{\Sigma}\right)$, and $g_{t\phi} = \frac{r_s r a \sin^2\theta}{\Sigma} $. In a physically realistic situation, the orbital angular momentum of a test particle and the spin angular momentum of a central rotating body need not necessarily be aligned. Here, for simplicity we have restricted our attention to the orbits in the equatorial plane ($\theta=\frac\pi2$).
Eqs. (\ref{conserved1}), (\ref{conserved2}) can be solved for $U^t$ and $U^{\phi}$, and we get,
\begin{equation}
    U^t =  \frac{1}{\Delta} \left[\left(r^2 + a^2 + \frac{r_s  a^2 }{r}\right) e - \left(\frac{r_s  a }{r}\right) L\right] \,\, ,
    \label{four0}
\end{equation}
\begin{equation}
    U^{\phi} = \frac{1}{\Delta} \left[\left(\frac{r_s a }{r}\right) e + \left(1 - \frac{r_s }{r}\right) L\right]\,\, .
    \label{four1}
\end{equation}

Using normalization condition $U^{\alpha}U_{\alpha}=-1$ of timelike geodesics and also using the Eqs. (\ref{four0}), (\ref{four1}), we can derive $r$-component of the four velocity $U^r$ as,
\begin{equation}
   U^r = \pm \sqrt{(e^2 - 1) + \frac{r_s}{r} - \frac{L^2 - a^2 (e^2-1)}{r^2} + \frac{r_s (L - a e)^2}{r^3}}\,\, .
   \label{four2}
\end{equation}

    \begin{figure*}
	\includegraphics[scale=0.42]{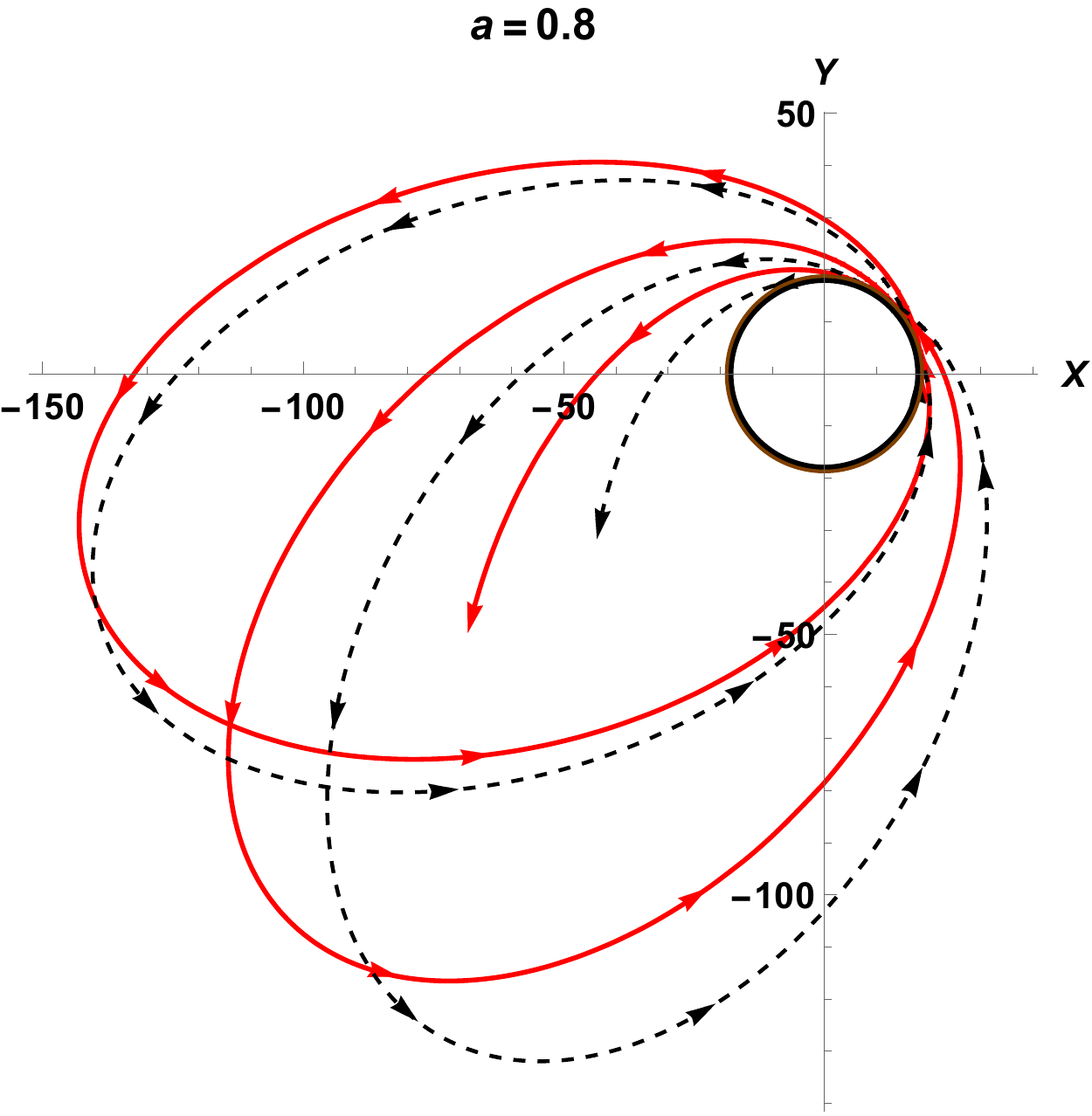}
	\includegraphics[scale=0.42]{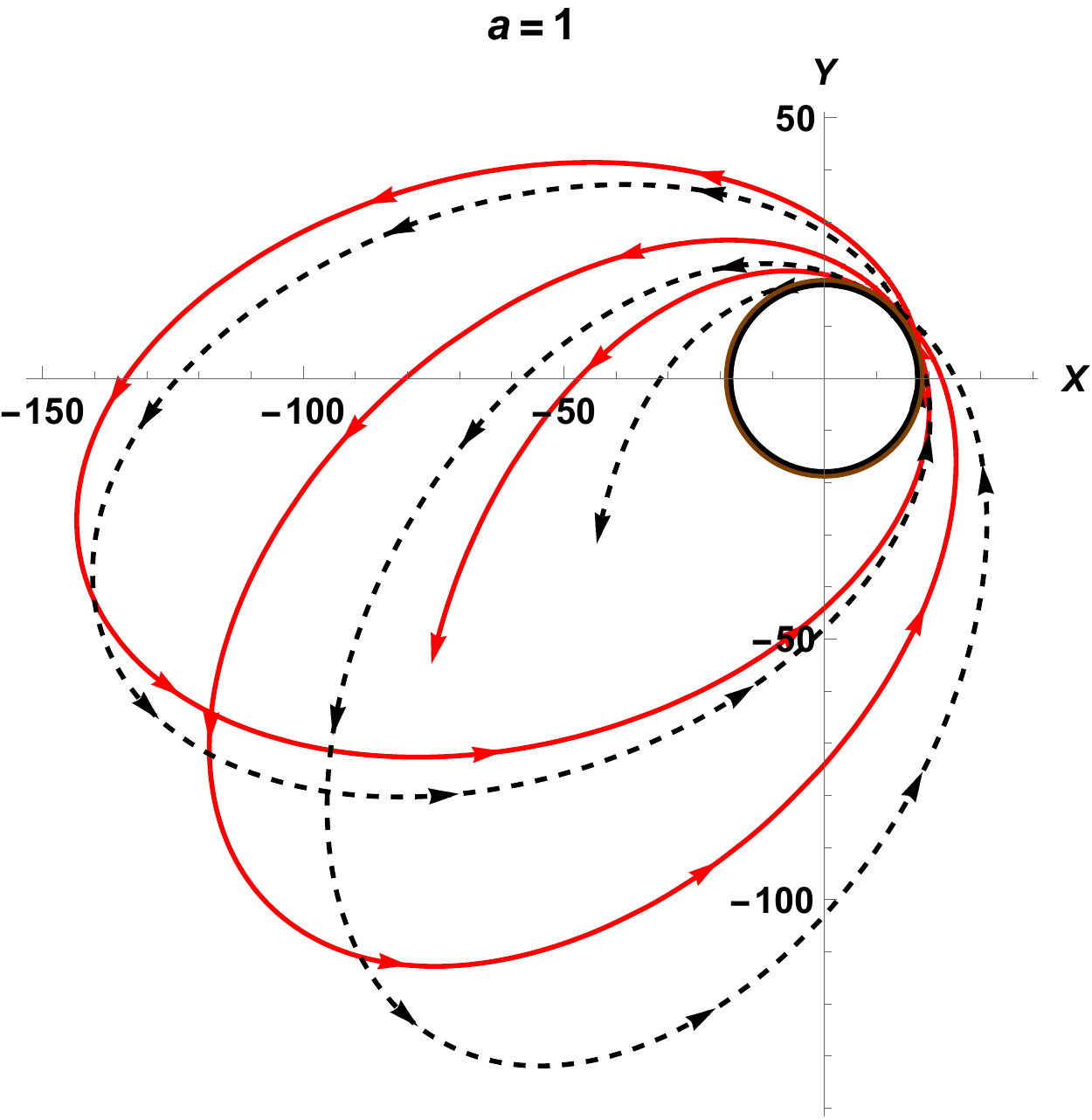}
	\includegraphics[scale=0.42]{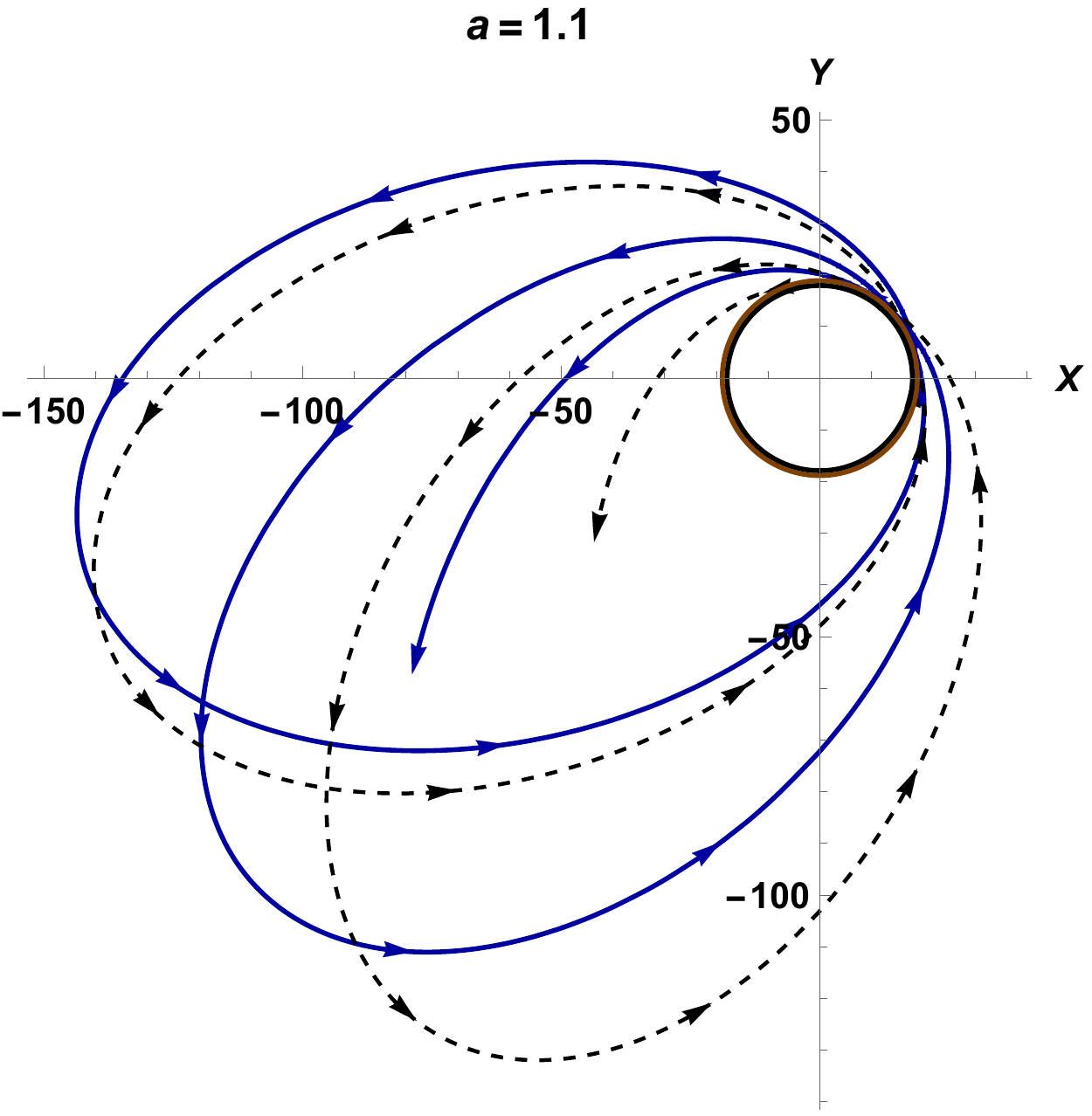}
	\includegraphics[scale=0.42]{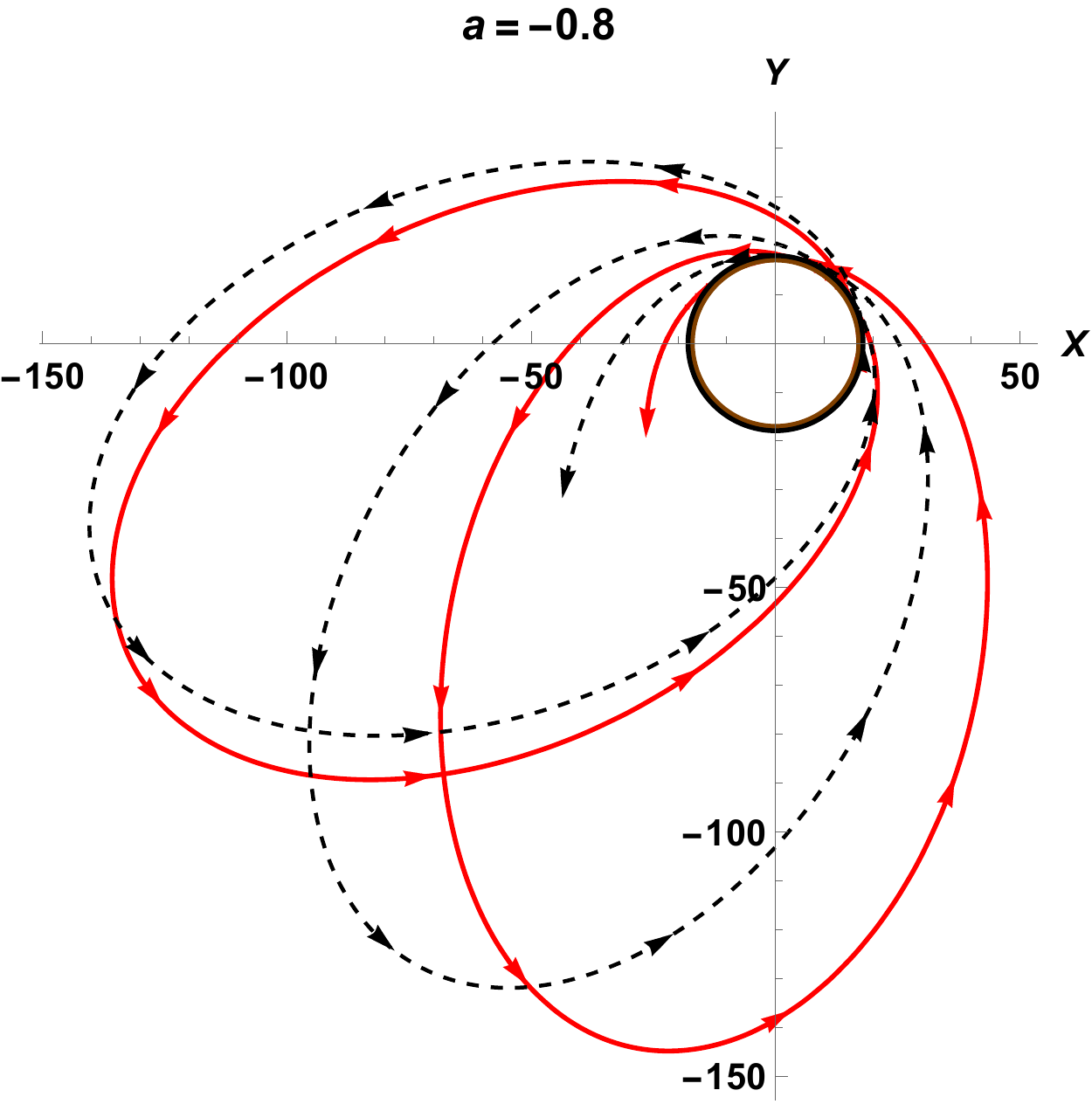}
	\includegraphics[scale=0.42]{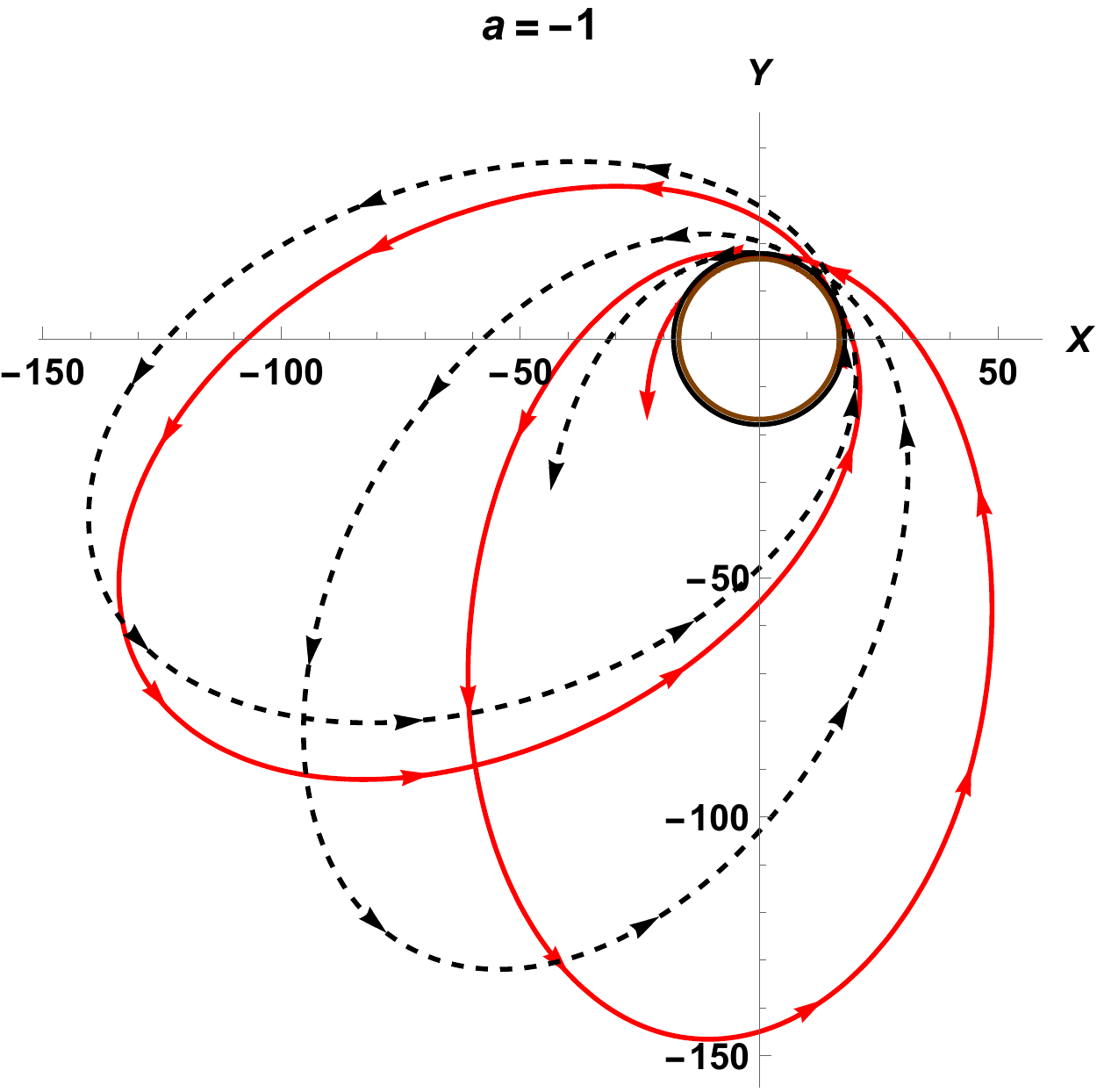}
	\includegraphics[scale=0.42]{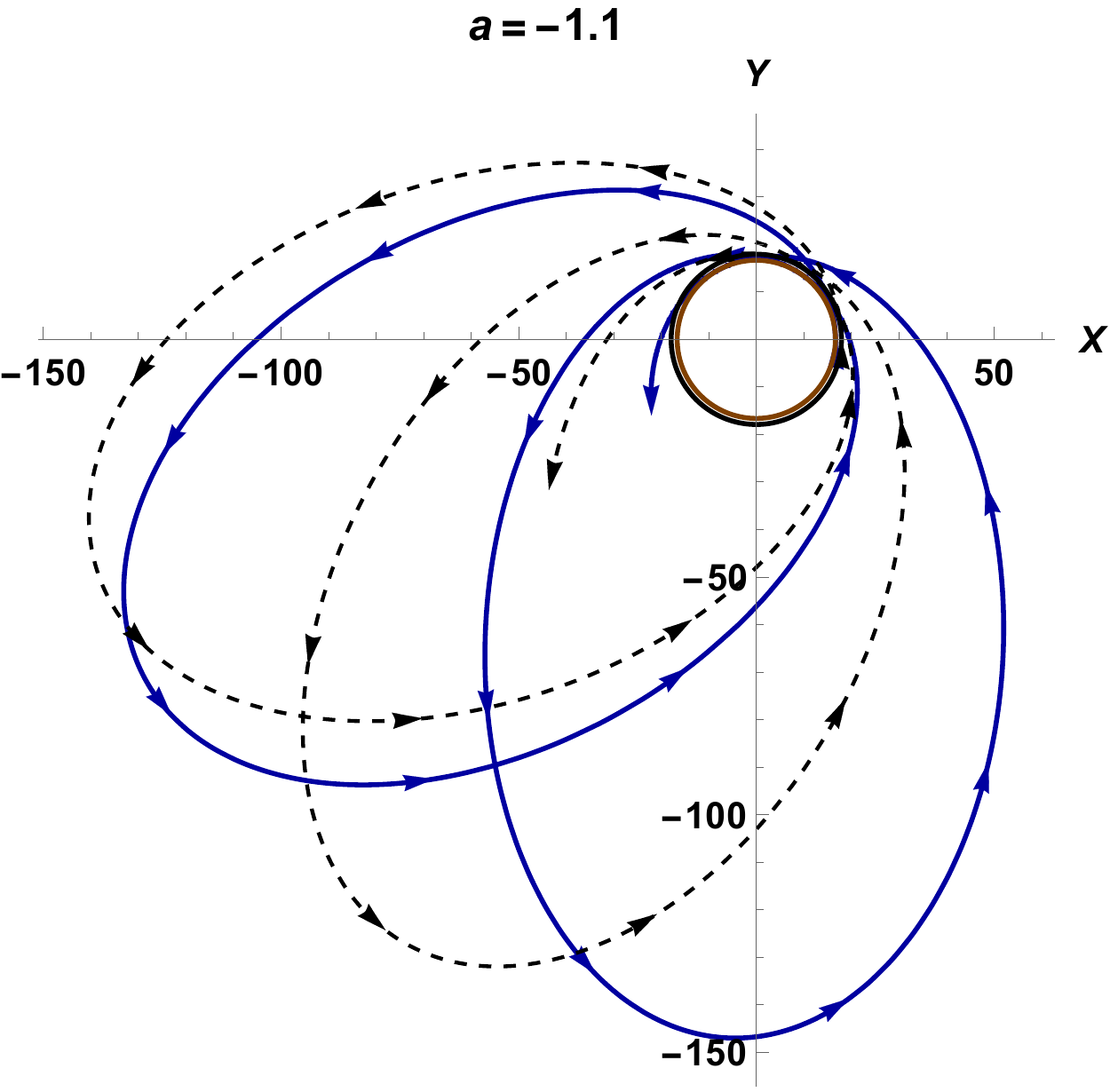}
\caption{Above figures show orbits of a test particle in the Kerr spacetime. Red solid lines indicate the particle's orbits in the Kerr black hole for $a=0.8, r_{min}=18.62$ and $a=-0.8, r_{min}=17.03$ (first column). In the  extreme Kerr black hole, for $a=1, r_{min}=18.78$, and for $a=-1$,  we have $r_{min}=16.77$ (second column). The  blue solid lines indicate particle's orbits in the Kerr naked singularity for $a=1.1, r_{min}=18.86$, and $a=-1.1, r_{min}=16.64$ (third column). Black dotted lines indicate particle's orbits in the Schwarzschild spacetime (for a=0, $r_{min}=17.91$). Brown circles represent the minimum approach of the particle towards the center. Here, we have considered $M=1$, $L=6$ and total energy is $E=-0.006$. }
	\label{orbit2}
    \end{figure*}
Here, $\pm$ signatures are corresponding to the radially outgoing and incoming timelike geodesics respectively. The expression, in Eq.~(\ref{four2}), is equivalent to the kinetic energy of a test particle. The total relativistic energy is defined as,
\begin{equation}
    E = \frac{1}{2} (e^2 - 1) = \frac{1}{2} (U^r)^2 + V_{eff}(r)\,\, .\label{E1}
\end{equation}
Using the expression of $U^r$ (Eq.~(\ref{four2})) and the expression of total relativistic energy (Eq.~(\ref{E1})), we get the following expression of effective potential,
\begin{equation}
    V_{eff}(r) = -\frac{r_s}{2r} + \frac{L^2 - a^2 (e^2-1)}{2r^2} - \frac{r_s (L-a e)^2}{2 r^3}\,\, .
    \label{veff1}
\end{equation}
Above expression of the effective potential is only applicable for equatorial timelike geodesics. For bound orbits, the total energy of the particle is greater than or equal to the minimum effective potential. The minimum effective potential is determined as,
\begin{equation}
    \frac{dV_{eff}}{dr}|_{r_b} = 0 ;        \frac{d^2 V_{eff}}{dr^2}|_{r_b}>0\,\, ,
    \label{con1}
\end{equation}
where the effective potential has a minimum at $r=r_b$.
Using Eq.~(\ref{veff1}) and Eq.~(\ref{con1}), we get the following expression of $r_b$, 
\begin{widetext}
\begin{equation}
    r_b = \frac{1}{2M}\big(L^2 + a^2 (1 - e^2) + \sqrt{(L^2 + a^2 (1 - e^2))^2 - 12M^2 (L - ae)^2}\big)\,\, .
    \label{rb1}
\end{equation}
The minimum effective potential at $r = r_b$ is,
\begin{multline}
    V_{eff}(r_b) = \frac{-1}{(L^2 + a^2 (1 - e^2) + \sqrt{(L^2 + a^2(1 - e^2))^2 - 12M^2 (L - a  e)^2})^3}(2M^2 (a^4 (1 - e^2)^2 + 16a L e M^2
    \\
    + L^2(L^2 - 8M^2 + \sqrt{(L^2 + a^2(1 - e^2))^2 - 12M^2 (L - a e)^2}) + a^2 (2L^2 (1 - e^2) 
    \\
    - 8M^2 + (1 - e^2) \sqrt{(L^2 + a^2(1 - e^2))^2 - 12M^2 (L - a e)^2})))\,\, .
\end{multline}
\end{widetext}
The bound orbits exist for $V_{min}\leq E<0$. Using the bound orbit conditions, we can determine the shape of the orbits, that gives how $r$ changes in the equatorial plane with respect to $\phi$,
\begin{equation}
    \frac{dr}{d\phi} = \frac{\pm \sqrt{(e^2 - 1) + \frac{2 M}{r} - \frac{L^2 - a^2 (e^2-1)}{r^2} + \frac{2 M (L - a e)^2}{r^3}}}{\frac{1}{\Delta} \left[\left(\frac{r_s a }{r}\right) e + \left(1 - \frac{r_s }{r}\right) L\right]}\,\, .
    \label{shapeorbits}
\end{equation}
Using Eq.~(\ref{shapeorbits}), we can derive second order differential orbit equation of a massive test particle in Kerr spacetime,
\begin{widetext}
\begin{multline}
    \frac{d^2u}{d\phi^2}= \frac{1 - 2Mu + a^2 u^2}{(L - 2Mu (L - a e))^3} \left[M (L - 2a e (e^2 - 1)) + u \left(L (-L^2 + 3a^2 (e^2 - 1)) - 2M^2 (2L + 2a e) + u B(\phi)\right) \right]\,\, ,
    \label{orbiteq1}
\end{multline}
where $u = \frac{1}{r}$ and 
\begin{multline}
    B(\phi) = M (7L^3 - 6a e L^2 + a^2L (11 - 3e^2) + 2a^3e (e^2-1)) + 4M^3 (L - a e) + u [-3a^2L^3 + 3a^4L (e^2 - 1)
    \\
    + 2M^2 (L - a e)(-8L^2 + a (7Le + a (e^2-5))) - Mu (L - a e) (a^2(-11L^2 + a (7Le + 4a (e^2 - 1)))
    \\
    - 12M^2(L - a e)^2 + 10a^2Mu(L-a e)^2)  ]\,\,\nonumber .
\end{multline}
\end{widetext}

We numerically solve the above orbit equation (Eq.~(\ref{orbiteq1})) to investigate the nature and shape of bound orbits of a test particle which is freely falling in Kerr spacetime. Fig.~(\ref{orbit1}) shows the bound orbits of a test particle in the Kerr black hole spacetime and Kerr naked singularity spacetime. In that figure, we show timelike bound orbits for spin parameters $a=\pm 0.8, \pm 1, \pm 1.1$. As we know, the values of spin parameter $a=\pm 0.8, \pm 1, \pm 1.1$ correspond to the Kerr black hole, extreme Kerr black hole and Kerr naked singularity respectively. In Fig.~(\ref{orbit1}), we consider the particle's total energy $E=-0.001$, angular momentum $L=12$ and the mass of the black hole to be $M=1$. The orbit shown by black dotted lines represents the timelike orbits in Schwarzschild spacetime (i.e. $a=0$). In Fig.(\ref{orbit1}), the timelike bound orbits in Kerr black hole spacetime (i.e. $a<1$), and in Kerr naked singularity spacetime (i.e. $a>1$) are shown by solid red lines and solid blue lines respectively.  It can be seen that the orbital precession in Kerr spacetime is distinguishable from the orbital precession in the Schwarzschild spacetime. 

All the orbits in Fig.~(\ref{orbit1}) show a positive precession and the non-zero spin parameter changes the minimum approach ($r_{min}$) and perihelion shift of those orbits. One can see that for $a>0$, the minimum approach of the particle (Periastron point) increases as the value of spin parameter increases. On the other hand, for $a<0$, the minimum approach of the particle  decreases as the value of spin parameter decreases. This effect of spin parameter can also be seen in Fig.~(\ref{orbit2}), where the particle's total energy $E=-0.006$ and angular momentum $L=6$. Since the angular momentum ($L$) considered in Fig.(\ref{orbit2}) is smaller than that of the Fig.(\ref{orbit1}), the minimum approaches of the orbits in the Fig.~(\ref{orbit2}) are much smaller than the minimum approaches of the orbits in the Fig.~(\ref{orbit1}). Therefore, the frame-dragging effect of Kerr black hole geometry is much higher in the second case (i.e. for $L=6$). However, it can be verified that all the orbits in Fig.~(\ref{orbit2}) always have a positive precession.

There exists a radial limit under which no stable bound orbit is possible in Kerr spacetime. As we know, in Schwarzschild spacetime, there exists a minimum value of the radius of the stable circular orbit and it is known as the radius ($r_{ISCO}$) of innermost stable circular orbit (ISCO). For Schwarzschild spacetime, the ISCO is at $r_{ISCO}=6M$. If we put a condition that the $r_b$ in  Eq.~(\ref{rb1}) should always be real, then the $r_{ISCO}$ for Kerr metric can be written as,
\begin{eqnarray}
    r_{ISCO} &=& 3M - \sqrt{3}ae + \sqrt{9M^2-6\sqrt{3}aeM-3a^2(1-e^2)}\nonumber\,\, ,\\
    &\approx& 6M - 2\sqrt{3}a\bigg(e+\dfrac{a}{4\sqrt{3}M} \bigg)\,\, ,
    \label{rISCO}
\end{eqnarray}
where the first expression is the exact expression of $r_{ISCO}$ and the second expression is the approximate one, where we consider small values of the spin parameter. Using the first expression $r_{ISCO}$, one can verify that the value of $r_{ISCO}$ is real and finite when $a<M$ (i.e.  Kerr black hole). However, when $a>M$ (i.e.  Kerr naked singularity) there exist no real value of $r_{ISCO}$, which implies that stable circular orbits can extend up to the singularity.
The second expression is useful to understand how much the ISCO radius differs from $6M$ (i.e. the ISCO radius in Schwarzschild spacetime) due to the non-zero value of spin parameter. Any bound orbits with a minimum approach ($r_{min}$) close to $r_{ISCO}$, can have a very large perihelion shift.
One can derive the expression for the smallest possible value of $r_{min}$ ($r_{min0}$) of a bound timelike orbit in Kerr spacetime by finding the solution of $V_{eff}|_{r_{min0}}=0$, $\frac{dV_{eff}}{dr}|_{r_{min0}}=0$ and $\frac{d^2 V_{eff}}{dr^2}|_{r_{min0}}<0$. The expression of $r_{min0}$ can be written as,
\begin{equation}
r_{min0}\approx 4M-a\left(2e+\frac{a}{4M}\right)\,\, .
\end{equation}
Timelike bound orbits with the above value of minimum approach have large amount of perihelion shift. One can verify that though the perihelion shift is large near $r_{min0}$, the precession is always positive.
However, this large perihelion shift of timelike bound orbits cannot be seen or verified in the stellar motions of `S' stars. For example, S2 star has a minimum approach which is $r_{min}= 2800M$. Therefore, if we consider the central body (Sgr-A*) as a Kerr black hole, then the perihelion shift of the orbit of S2 would have a very small positive value, and that would be very close to the expected value of perihelion shift of the S2 star in a Schwarzschild background. Hence, the orbits of S2 star should always have a positive precession if we consider Schwarzschild or Kerr black hole at the center of our galaxy (Sgr-A*).  

In Fig.~(\ref{orbit1}) and Fig.~(\ref{orbit2}), we show the positive precession of timelike bound orbits (on $\theta=\frac\pi2$ plane) in Kerr spacetime for some particular values of $L$, $E$ and $a$. However, by those figures we cannot or do not actually claim to prove that the phenomenon of negative precession is always forbidden or absent in Kerr spacetimes. To prove that for timelike bound orbits in Kerr spacetime, we need to analytically solve the orbit equation (Eq.~(\ref{orbiteq1})). However, as the Eq.~(\ref{orbiteq1}) is a fairly complicated non-linear differential equation, we attempt to  solve it here either numerically, or using a suitable approximation. Therefore, in the next section, we investigate the nature of perihelion shift of the orbits in Kerr spacetime using an approximation technique. 
\begin{figure*}
\centering
\subfigure[$M=1$, $~3.8\leq L\leq 5$, $~0.895\leq e\leq 0.999$]
{\includegraphics[scale=0.5]{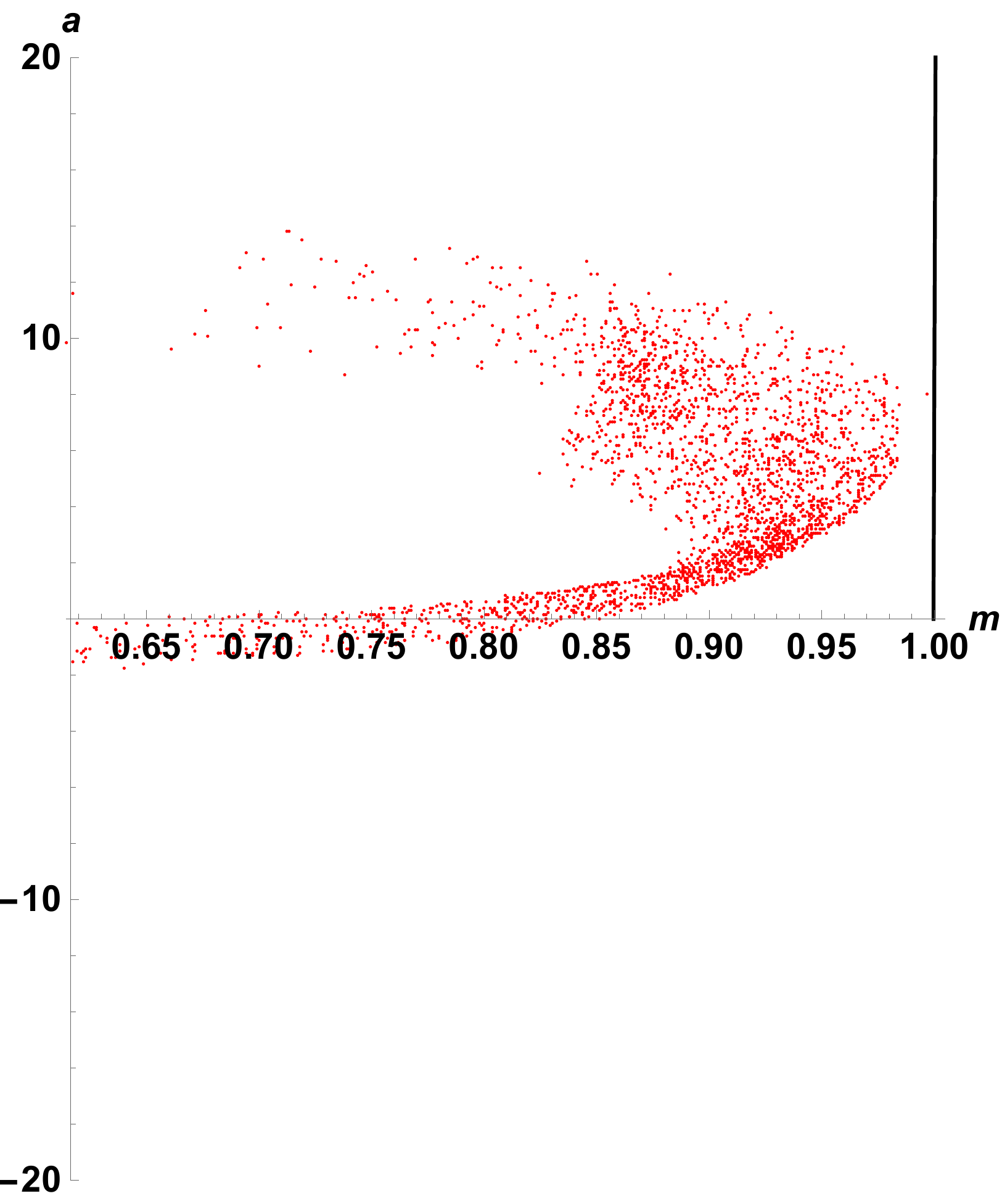}\label{re1}}
\hspace{0.1cm}
\subfigure[$M=1$, $~5\leq L\leq 10$, $~0.895\leq e\leq 0.999$]
{\includegraphics[scale=0.5]{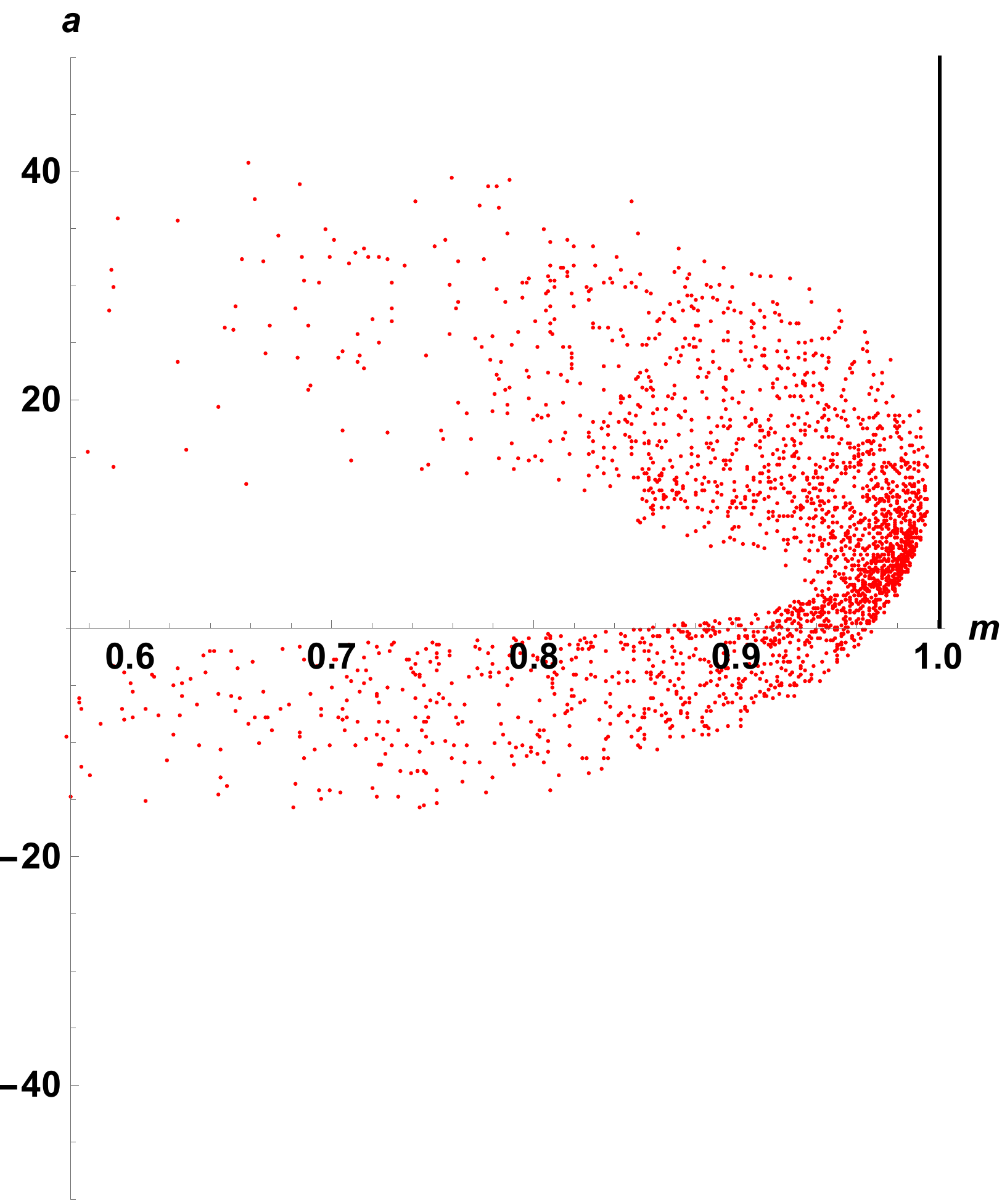}\label{re2}}
\hspace{0.1cm}
\subfigure[$M=1$, $~10\leq L\leq 20$, $~0.895\leq e\leq 0.999$]
{\includegraphics[scale=0.5]{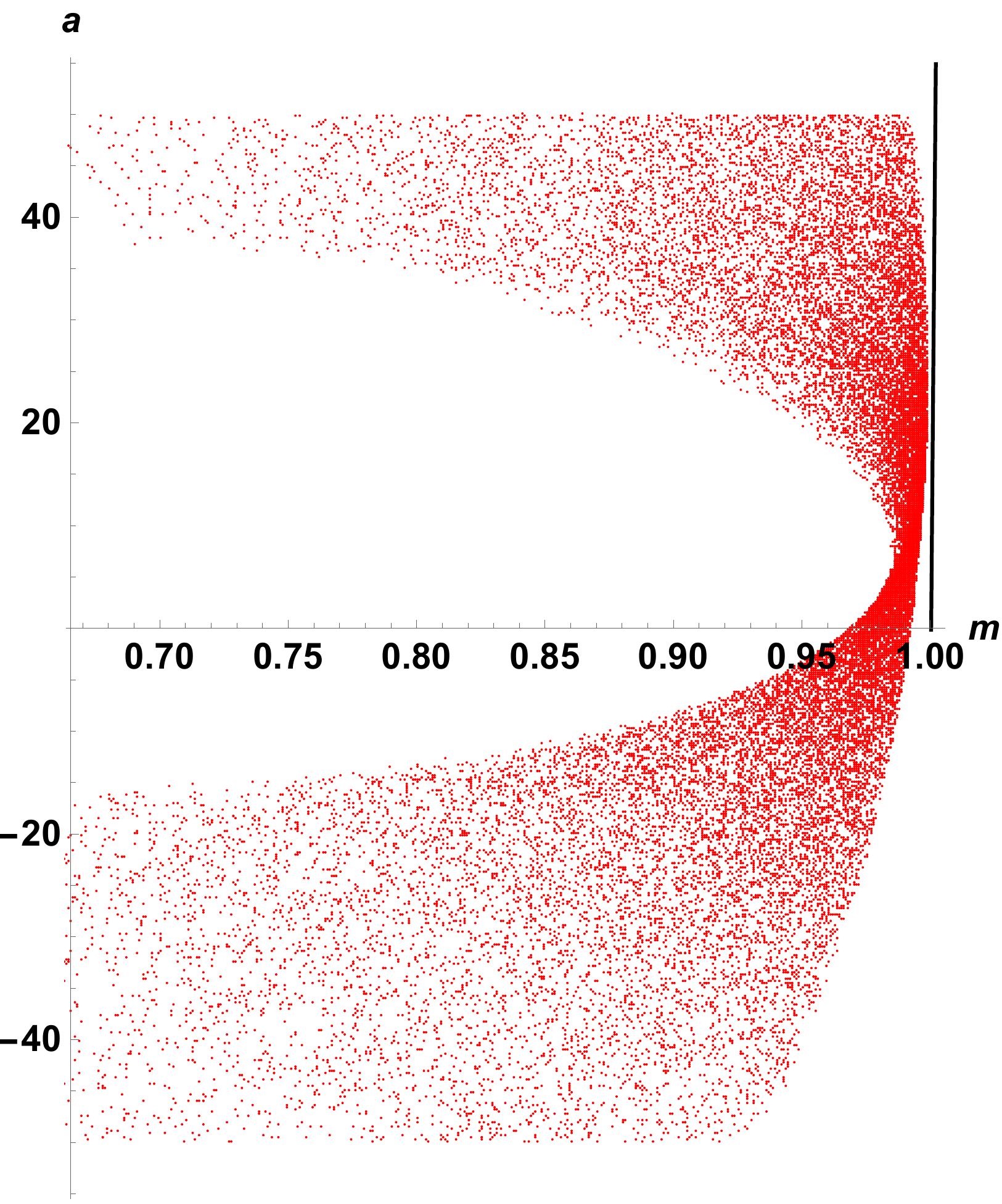}\label{re3}}
\hspace{0.1cm}
\subfigure[$M=1$, $~20\leq L\leq 35$, $~0.895\leq e\leq 0.999$]
{\includegraphics[scale=0.5]{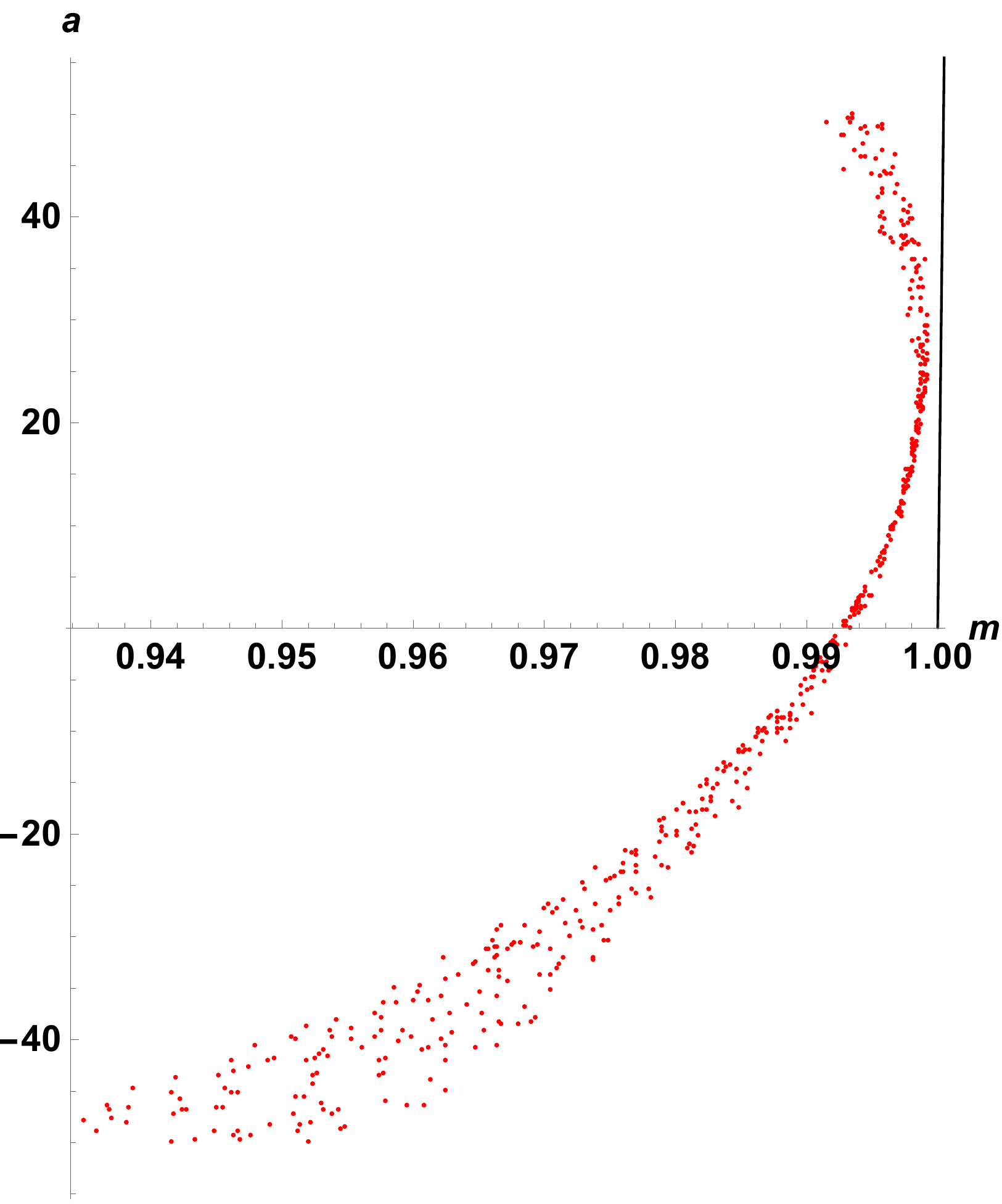}\label{re4}}
 \caption{The above figure shows the solution points for the timelike bound orbits in $m$ and $a$ coordinates. We have 
 $0< m<1$ and $m>1$, that represent the positive precession and negative precession of timelike bound orbits respectively.}
 \label{mvsa}
\end{figure*}

\section{An approximate solution of orbit equation}\label{three}
In this section, we investigate whether a negative precession of timelike orbits of particles is possible for any value of $L,e$ and $a$. For this purpose, we use an approximate method here where we only consider low eccentric orbits. Therefore, for the approximate solution of the Eq.~(\ref{orbiteq1}), we only consider upto the first order expression in eccentricity $\epsilon$ \cite{Parth, Struck:2005hi,Struck:2005oi}. The approximate solution can be written as, 
\begin{equation}
u(\phi)=\frac{1}{M p}(1+\epsilon\cos(m\phi)+\mathcal{O}(\epsilon^2))\,\, .
\label{18}
\end{equation}
where $p$ and $m$ are real positive constants and $m>1$ represents the precession of the timelike bound orbits in negative direction, while $m<1$ represents the precession in positive direction. 
When we substitute the above expression of $u(\phi)$ in the orbit equation (Eq.~(\ref{orbiteq1})) and separate the zeroth order terms and the first order terms of $\epsilon$, we get an expression of $m$ in terms of $p$. From the zeroth order terms, we get an equation of fifth order polynomial of $p$,
\begin{eqnarray}
    &g_5&(L,a,e,M) p^5 + g_4(L,a,e,M) p^4 + g_3(L,a,e,M) p^3\nonumber \\&+& g_2(L,a,e,M) p^2 + g_1(L,a,e,M) p + g_0(L,a,e,M) = 0\,\, ,\nonumber\\ 
    \label{polynomial}
\end{eqnarray}
where
\begin{eqnarray}
    g_5(L,a,e,M) &=& -(2ae(1-e^2)+L)M^4\,\, ,\nonumber\\
    g_4(L,a,e,M) &=& (3a^2(1-e^2)L+L^3+2aeM^2+4LM^2)M^2\,\, ,\nonumber\\
    g_3(L,a,e,M) &=& (2a^3e(1-e^2)-a^2(11-3e^2)L\nonumber\\&-&7L^3-4LM^2+2ae(3L^2+2M^2) ) M^2\,\, ,\nonumber\\
    g_2(L,a,e,M) &=& 3a^4L(1-e^2)-2ea^3M^2(5-e^2)-30aeL^2M^2\nonumber\\&+&16L^3M^2+a^2L(3L^2+2M^2(5+6e^2))\,\, ,\nonumber\\
    g_1(L,a,e,M) &=& -a^2(ae-L) (-4a^2(1-e^2)+7aeL-11L^2)\nonumber\\&+&12M^2(ae-L)^3\,\, ,\nonumber\\
    g_0(L,a,e,M) &=& -10a^2 (ae-L)^3\,\, .\nonumber
\end{eqnarray}
It is very difficult to get an analytical solution of the above fifth order polynomial equation (Eq.~(\ref{polynomial})). Therefore, we can use numerical technique and get five solutions of $p$. One can verify that among those five solutions only one solution has real and positive value. Now, we get the following expression of $m$ from the first order term of $\epsilon$, 
\begin{widetext}
\begin{multline}
    m^2 = \dfrac{-1}{M^4 (2ae+L(p-2))^4 p^3} \big(f_7(L,a,e,M) p^7 + f_6(L,a,e,M) p^6 + f_5(L,a,e,M) p^5 + f_4(L,a,e,M) p^4 \\
    + f_3(L,a,e,M) p^3 + f_2(L,a,e,M) p^2 + f_1(L,a,e,M) p + f_0(L,a,e,M)\big)\,\, ,
    \label{msquare}
\end{multline}
where\\

\begin{align*}
  & f_7(L,a,e,M) =  M^4 (-L^4 - 8 a e^3 L M^2 + 3 a^2 (-1 + e^2) (L^2 + 4 e^2 M^2))\,\, ,\\
  & f_6(L,a,e,M) = -2 M^4 (6 a^3 e (-1 + e^2) L - 7 L^4 + 4 a e L (L^2 - 2 e^2 M^2) + a^2 (3 (-4 + e^2) L^2 + 4 e^2 (-3 + 2 e^2) M^2))\,\, ,\\
  & f_5(L,a,e,M) = -6 L^2 M^2 (-3 a^4 (-1 + e^2) - 16 a e L M^2 + 12 L^2 M^2 + 2 a^2 (L^2 + 2 (3 + e^2) M^2))\,\, ,\\
  &  f_4(L,a,e,M) = 4 L M^2 (6 a^5 e (-1 + e^2) + 3 a^4 (7 - 2 e^2) L - 96 a e L^2 M^2 + 44 L^3 M^2 - 8 a^3 e (3 L^2 + (3 + e^2) M^2)\\
  &\hspace*{3cm} + 12 a^2 L (2 L^2 + (2 + 5 e^2) M^2))\,\, ,\\
& f_3(L,a,e,M) = 15 a^6 (-1 + e^2) L^2 - 15 a^4 L^4 + 12 a^2 (a e - L) (a^3 e (-1 + e^2) - a^2 (-11 + e^2) L\\
  &\hspace*{3cm}- 24 a e L^2 + 24 L^3) M^2 - 16 (-a e + L)^2 (a^2 (3 + e^2) - 14 a e L + 13 L^2) M^4\,\, ,\\
  & f_2(L,a,e,M) = 6 a^4 (a e - L) L (6 a^2 (-1 + e^2) + 7 a e L - 13 L^2) + 24 a^2 (-a e + L)^2 (3 a^2- 16 a e L + 16 L^2) M^2\\
  &\hspace*{3cm}+ 96 (-a e + L)^4 M^4\,\, ,\\
  & f_1(L,a,e,M) = 8 a^4 (-a e + L)^2 (3 a^2 (-1 + e^2) + 14 a e L - 17 L^2) - 192 a^2 (-a e + L)^4 M^2\,\, ,\\
  & f_0(L,a,e,M) = 80 a^4 (-a e + L)^4\,\, .
\end{align*}
\end{widetext}
We put the numerical solutions of $p$ in the expression of $m$ given in Eq.~(\ref{msquare}) and get the numerical values of $m$. As it is mentioned before, $m>1$ implies negative precession of the timelike orbits in Kerr spacetime. Therefore, we verify whether there exist any parameter space regions where $m$ is greater than one. To get numerical solution of $p$, we consider some specific physically realistic parameter spaces for the parameters $a,L$ and $e$. In our numerical analysis, we always consider $M=1$, and $a$ varies from $-40$ to $40$, $L$ varies from $3.8$ to $35$ and $e$ varies from $0.895$ to $0.999$. The spin parameters $a<-40$ and $a>40$ are not physically realistic as they may represent somewhat extreme spin situations. When specific energy $e<0.8$, solutions for bound stable orbits becomes hard to find out. Specific energy cannot have values beyond one, since the total energy of particle becomes positive, which implies unbound orbits. Our approximate solution mentioned in Eq.~(\ref{18}) is not a good approximation for the orbits of high eccentricity. For $L$ less than $3.8$, the eccentricity of the orbits becomes very high and therefore, using our approximation method, we cannot do numerical analysis in $L<3.8$ region.

In Fig.~(\ref{mvsa}), we show the solution points for which bound stable orbits are possible. We show those points in the $m$ and $a$ coordinates. We know that $m$ is a function of $a$, $L$ and $e$. Therefore, for a particular value of $a$, we can get many values for $m$, since at every points other two variables ($L$ and $e$) are varying. We separate $3.8\leq L\leq 35$ into four sections so that we can get more data points in every section. From Figs.~(\ref{re1}, \ref{re2}, \ref{re3}, \ref{re4}), we can see that all the solution points are inside the region $0\leq m\leq 1$ which indicates that the precession of orbits in Kerr spacetime is always positive for the classes we have considered. One can see from Fig.~(\ref{mvsa}) that for any absolute value of the spin parameter, the positive perihelion shift becomes larger when the sign of the spin changes from positive to negative. This phenomenon also can be seen from Figs.~(\ref{orbit1},\ref{orbit2}). Therefore, if negative precession is not possible for any positive value of spin parameter then it would not be possible for any negative value of spin parameter either.  One can consider larger interval of $a$ to verify whether negative precession of orbits is possible for larger values of $a$ (i.e. $a>50$), or smaller values of $a$ (i.e. $a<-50$). However, it can be verified from the Fig.~(\ref{mvsa}) that the solution points are very close to $m=1$ in the interval $0\leq a\leq 30$ and after that range, those points are diverging away from $m=1$.  Therefore, from the above analysis it can be stated that the negative precession is forbidden under the given approximation in the Kerr spacetime. 

Till now, in this section, we have done all the analysis considering only the low eccentricity approximation and we show that negative precession of the orbits of a test particle is not possible in Kerr spacetime for the equatorial orbits. We do not consider any weak field approximation for our numerical analysis.
Since eccentricity of a orbit is not directly related with the perihelion distance, a highly eccentric orbit can be far away from the center, whereas a low eccentric orbit can be very close to the center. Therefore, in our analysis, the orbit of the test particle can be very close or far away from the singularity.

One can consider weak field approximation along with the small eccentricity approximation to get analytical solutions of $p$ and $m$. From that expression of $m$, one can get the expression of $m$ in the Schwarzschild limit (i.e. $a\rightarrow 0$).     If we want to consider weak field approximation, we can neglect third and higher order powers of $u(\phi)$ in the expression of the orbit equation (Eq.~(\ref{orbiteq1})). The approximate orbit equation upto the second order power of $ u(\phi)$ is given by,
\begin{widetext}
\begin{multline}
    \frac{d^2u}{d\phi^2} = \frac{1}{L^5}[M L^2(L + 2a e (1-e^2)) - L (L^4 + 8a Le^3M^2 + 3a^2(1-e^2)(L^2 + 4e^2M^2))u + 3M (L^2(L^3+3a^2Le^2 
    \\
    + 6a^3e (1-e^2)) - 4a e^2(2e L^2 + 3a L(1-2e^2)-4a^2e(1-e^2))M^2)u^2].
    \label{weak}
\end{multline}
\end{widetext}
 
Substituting the expression of $u(\phi)$ (Eq.~(\ref{18})) in the above orbit equation (\ref{weak}), we get the following quadratic equation of $p$ from the coefficient of zeroth order power of eccentricity ($\epsilon$),
\begin{eqnarray}
\tilde{g}_2(L,a,e,M) p^2 +\tilde{g}_1(L,a,e,M) p + \tilde{g}_0(L,a,e,M) = 0\,\, ,\nonumber\\ 
    \label{polynomial2}
\end{eqnarray}
where
\begin{widetext}
\begin{eqnarray}
\tilde{g}_2(L,a,e,M)&=& 2a L^2 M^2 e (e^2 - 1) - L^3 M^2\nonumber\,\,,\\
\tilde{g}_1(L,a,e,M)&=& L\left[ L^4 + 8a L e^3 M^2 - 3a^2 (e^2 - 1)(L^2 + 4e^2 M^2)\right]\nonumber\,\,,\\
\tilde{g}_0(L,a,e,M)&=&[-3L^2(L^3 + 3L e^2 a^2 
    +6e a^3(1 - e^2)) + 12a e^2 M^2(2e L^2 + 3a L (1 - 2e^2) - 4e a^2(1-e^2))]\,\, .\nonumber
\end{eqnarray}
Now, we get the following expression of $m$ from the first order term of eccentricity ($\epsilon$), 
\begin{multline}
  m^2 = \frac{1}{p L^5}[96a^3 e^3 M^2 (e^2 - 1) + L^5 (p - 6) + 3a^2 L^3 (p - e^2 (p + 6))
+ 4a e L^2 (9a^2 (e^2 - 1) + 2e^2 M^2 (p + 6))
\\- 12L a^2 e^2 M^2 (-p - 6 + e^2 (p + 12))]\,\, .
\end{multline}
\end{widetext}
\begin{figure*}\centering
\subfigure[$~M=1$, $L=6$]
{\includegraphics[scale=0.450]{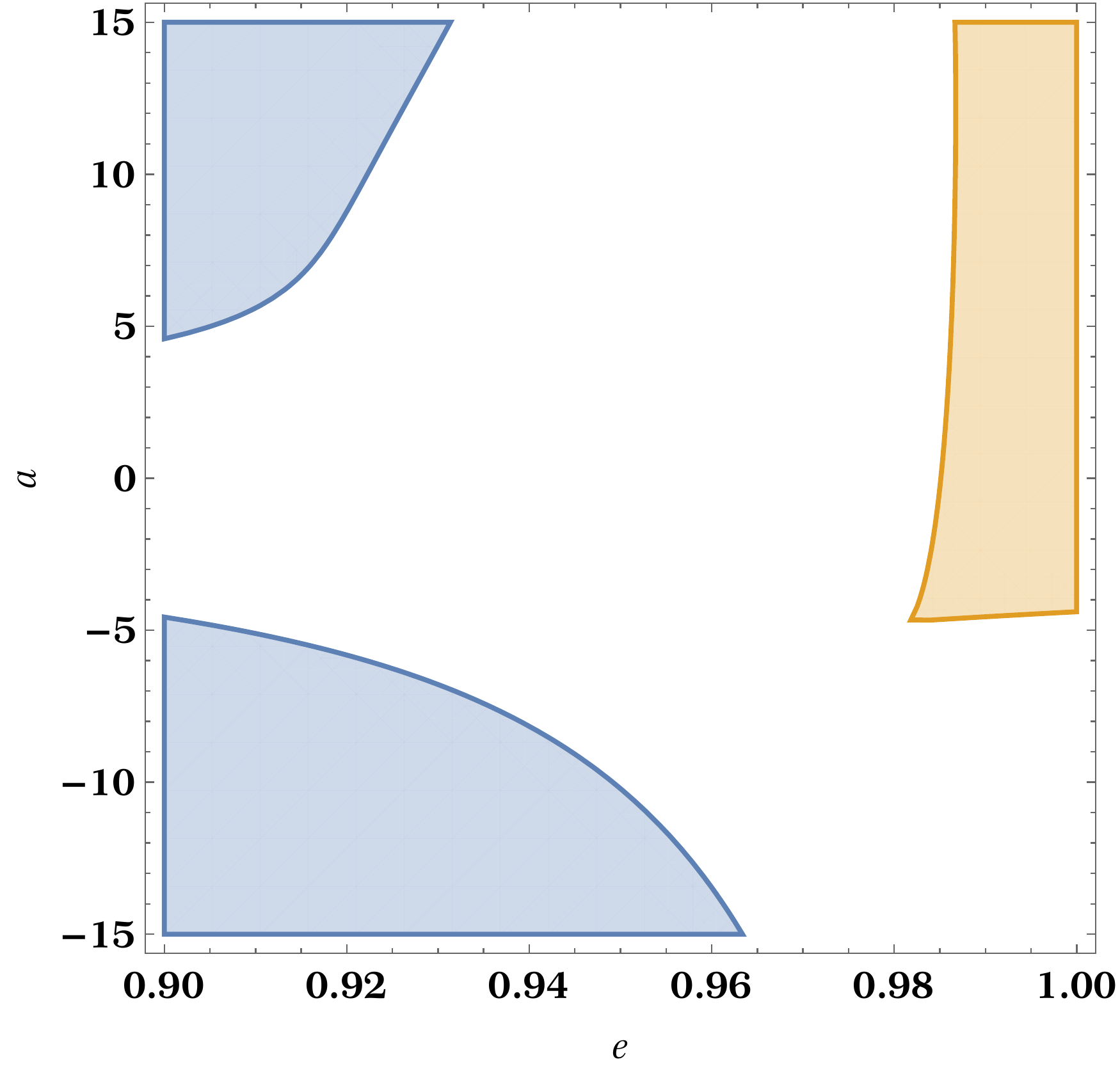}\label{figh6}}
\hspace{0.1cm}
\subfigure[$~M=1$, $L=12$]
{\includegraphics[scale=0.450]{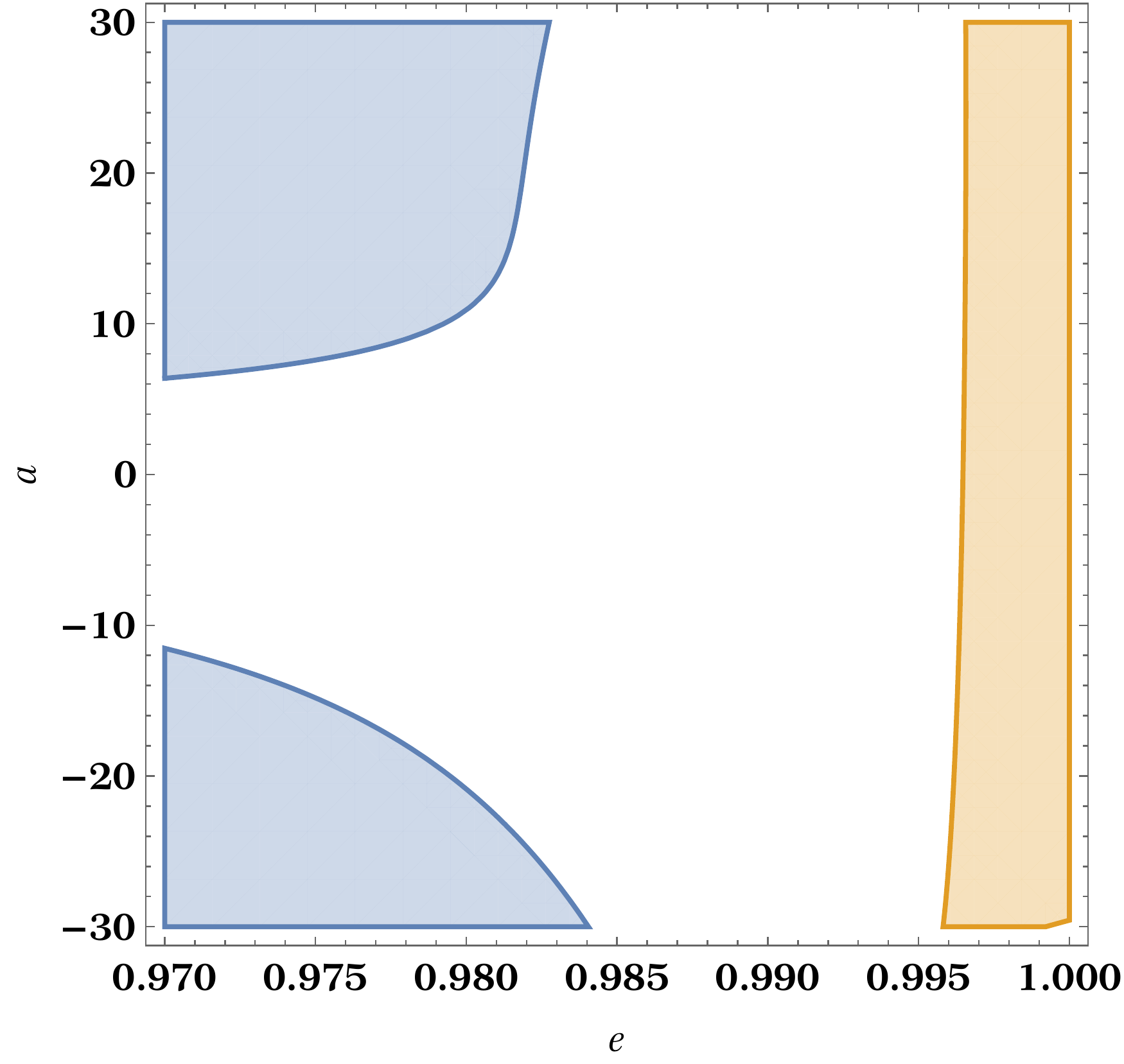}\label{figh12}}
	\caption{This figure shows region plots of $f>0$ and $V_{min}\leq E<0$ for M=1 and $L=6$~(Fig.~\ref{figh6}) and $L=12$ (Fig.~\ref{figh12}). Blue region represents region of negative precession ($f>0$) and orange region represents region of bound orbits ($V_{min}\leq E<0$). }
	\label{fig:3}
\end{figure*}
\begin{widetext}
One can solve the above quadratic equation analytically and get the following two roots of $p$,
\begin{multline}
    p_1 = \frac{1}{2L^2 M^2 (L + 2a e (1 - e^2))}[L^5 + 3a^2 L^3 (1 - e^2) + 8a L^2 M^2 e^3 + 12L a^2 e^2 M^2 (1 - e^2) 
    \\
    + \{L^2 (-12M^2 (L + 2a e (1 - e^2)) 
    (L^2 (L^3 + 3L a^2 e^2 + 6e a^3 (1 - e^2)) - 4a e^2 (2e L^2 + 3a L(1 - 2e^2)
    \\
    - 4e a^2(1 - e^2)) M^2) + (L^4 + 8a L e^3 M^2 + 3a^2 (1 - e^2)(L^2 + 4e^2 M^2))^2) \}^{1/2}]
\end{multline}

\begin{multline}
    p_2 = \frac{1}{2L^2 M^2 (L + 2a e (1 - e^2))} [L^5 + 3a^2 L^3 (1 - e^2) + 8a L^2 M^2 e^3 + 12L a^2 e^2 M^2 (1 - e^2)
    \\
    - \{L^2 (-12M^2 (L + 2a e (1 - e^2)) (L^2 (L^3 + 3L a^2 e^2 + 6e a^3 (1 - e^2)) - 4a e^2 (2e L^2 + 3a L(1 - 2e^2)
    \\
    - 4e a^2 (1 - e^2)) M^2) + (L^4 + 8a L e^3 M^2 + 3a^2 (1 - e^2)(L^2 + 4e^2 M^2))^2) \}^{1/2}]
\end{multline}
Now, it can be verified that for $p=p_2$, $m$ becomes imaginary and therefore, the real solution of the quadratic equation of $p$ is $p=p_1$. We get the following expression of $m$ after substituting $p=p_1$ in the expression of $m$,
\begin{equation}
m = (1 + f(a,M,L,e))^{1/4}\,\, ,
\label{m}
\end{equation}
where
\begin{multline}
    f(a,M,L,e) = \frac{1}{L^8}(-12M^2 L^6 + 8a e M^2 L^3 (L^2 (5e^2 - 3) + 12e^2 M^2) + 24 L e a^3 M^2 (e^2 - 1)(L^2 (e^2 + 3)
    \\
    + 4e^2 M^2(4e^2 - 1)) - 3a^4 (e^2 - 1)^2 (-3L^4 + 24L^2 e^2 M^2 + 80e^4 M^4) - 2a^2 L^2(3L^4 (e^2 - 1) 
    \\
    + 6L^2 e^2 M^2(1 + 2e^2) + 8e^2 M^4(8e^4 + 6e^2 - 9)))\,\, .
    \label{f}
\end{multline}
\end{widetext}
In the above expressions of $m$, $f>0$ implies $m>1$ which is the necessary condition for the negative precession and $f<0$ implies $m<1$ which is the necessary condition for the positive precession. As $m$ must be a real and positive number,  $f\ge -1$. Therefore, for the positive precession, we must have $-1<f<0$. In the Schwarzschild limit or with the approximation of negligibly small value of $a$, the expression of $m$ written above reduces to,
\begin{equation}
m = \left(1 -\dfrac{12 M^2}{L^2}\right)^{1/4}\,\, ,
\label{SCHm}
\end{equation}
where $f=\dfrac{-12 M^2}{L^2}$.
 Therefore, to satisfy the condition $-1<f<0$ we need,
\begin{equation}
\dfrac{L}{M}>\sqrt{12}\,\, .
\label{ratioLM}
\end{equation}
From the expression of $m$ in Eq.~(\ref{SCHm}), one can verify that in Schwarzschild spacetime, $m$ is always less than one which implies positive precession of timelike bound orbits \cite{Parth,Joshi:2019rdo}. If we consider $L>>M$, we can write down an approximate expression of $m$ for the Schwarzschild spacetime,
\begin{equation}
m = \left(1 -\dfrac{3 M^2}{L^2}\right)\,\, .
\end{equation}
From the above expression of $m$, we can get the following positive perihelion shift of timelike bound orbits in Schwarzschild spacetime \cite{Parth,Joshi:2019rdo},  
\begin{equation}
\delta\phi_{prec}=\frac{6\pi M^2}{L^2}\,\, .
\end{equation}
In Kerr spacetime, we can also get the minimum value of $\frac{L}{M} $ for bound orbits,
\begin{equation}
    \dfrac{L}{M} \approx \sqrt{12}-\dfrac{a}{M}\bigg(e+\dfrac{a}{\sqrt{12}M} \bigg)\,\, ,
    \label{KerrLMratio}
\end{equation}
where we consider only upto the second order power of $a$.

Previously, we have shown that the negative precession does not occur in Kerr spacetime, where we do not consider any weak field approximation. Now, with the weak field approximation, it should be obvious that we would get the same result. To verify that, in Fig.~(\ref{fig:3}), we show two region plots for fixed mass (M) and angular momentum (L), where we show the regions of the negative precession $f>0$ (i.e. the blue region) and the bound orbits $V_{min}\leq E<0$ (i.e. the orange region) for different values of $a$ and $e$. Existence of any common region between these two regions implies that in Kerr spacetime, negative precession of bound timelike orbit is possible. However, we do not find any of such overlapping regions. Therefore, again we get the same result that  we   have verified previously without considering any weak field approximation.

\section{Discussion and Conclusion}\label{four}
In this paper,  we have studied the nature of timelike bound orbits of a test particle in Kerr spacetime geometry. In order to find the particle trajectories in Kerr spacetime, we derive an orbit equation and solve the same numerically. In Fig.~(\ref{orbit1}), we show the trajectories of a test particle in Kerr spacetime for the spin parameter values $a=\pm 0.8, \pm 1, \pm 1.1$, where we take black hole mass $M=1$, specific angular momentum of the test particle $L=12$ and the total energy of the test particle $E=-0.001$. In Fig.~(\ref{orbit2}), we show particle trajectories for the same spin parameter values where $M=1, L=6$ and $ E=-0.006$. In both the cases we get positive precession of the bound orbits. We have investigated the particle trajectory only on the equatorial plane ($\theta=\pi/2$). 

Then in section (\ref{three}), we consider an approximation solution of the orbit equation (Eq.~(\ref{orbiteq1})), in order to understand the nature of perihelion shift of timelike bound orbits in Kerr spacetime. In that approximation, we consider only small values of eccentricity ($\epsilon$), and therefore, we neglect second and higher order terms of $\epsilon$. With this approximation, we show that in Kerr spacetime, negative precession of the timelike bound orbits is forbidden, no matter how much far or how much close the orbit is from the center.

Finally, we take a weak field approximation and we show that the solutions of Kerr spactime reduce to the solution of Schwarzschild with the approximation $a\rightarrow 0$. 
In this paper and in \cite{Parth,Joshi:2019rdo}, we show that negative precession of timelike orbits is not possible in Kerr and Schwarzschild spacetimes respectively. In \cite{Parth,Joshi:2019rdo}, we also showed that naked singularity models, such as JMN, JNW spacetimes admit both negative and positive precession of timelike orbits. 

As we know, GRAVITY and  SINFONI collaborations are continuously observing the stellar motions of `S' stars around the Milky-way galaxy center Sgr-A*. Hence, any evidence of negative precession of any `S' star can raise big question on the existence of Kerr black hole at the Milky-way galaxy center.  

Of course, we have not scanned here the full space of bound orbits around the Kerr black hole or the Kerr naked singularity. Our analysis, however, clearly points that in the classes of bound orbits we analysed, using the approximation and numerical techniques as we have stated, we have not found any negative precession for the bound orbits in both these cases. In particular, the case of Kerr black hole needs to be analysed in more detail to make sure if it forbids the negative precession for the bound timelike trajectories always. If that turns out to be the case, that will support the conjecture that black holes never allow for the negative precession, however, naked singularities allow the same as shown by some of the naked singularity spacetimes investigated as we pointed out here.


\end{document}